\documentstyle[12pt,aaspp4,epsfig]{article}
\lefthead{MACFADYEN, WOOSLEY, AND HEGER}
\righthead{}
\def\gtaprx {\lower .1ex\hbox{\rlap{\raise .6ex\hbox{\hskip .3ex
	{\ifmmode{\scriptscriptstyle >}\else
		{$\scriptscriptstyle >$}\fi}}}
	\kern -.4ex{\ifmmode{\scriptscriptstyle \sim}\else
		{$\scriptscriptstyle\sim$}\fi}}}
\def\ltaprx {\lower .1ex\hbox{\rlap{\raise .6ex\hbox{\hskip .3ex
	{\ifmmode{\scriptscriptstyle <}\else
		{$\scriptscriptstyle <$}\fi}}}
	\kern -.4ex{\ifmmode{\scriptscriptstyle \sim}\else
		{$\scriptscriptstyle\sim$}\fi}}}
\def \sun {$_{\scriptscriptstyle \odot}$}
\def \sune {_{\scriptscriptstyle \odot}}

\newcommand{\g}{{\mathrm{g}}} 
\newcommand{\cm}{{\mathrm{cm}}} 
\newcommand{\km}{{\mathrm{km}}} 
\newcommand{\Msun}{{\mathrm{M}_{\odot}}}
\newcommand{\Sec}{{\mathrm{s}}}
\newcommand{\erg}{{\mathrm{erg}}} 
 
\newcommand{\Ep}[1]{{10^{#1}}}

\newcommand{\foe}{{\Ep{51}\,\erg}}
\newcommand{\Foe}{{\cdot\foe}}
\newcommand{\aP}{{\alpha_{\mathrm{pist}}}}

\newcommand{\lSect}[1]{{\label{sec:#1}}}
\newcommand{\lFig}[1]{{\label{fig:#1}}}

\newcommand{\lTab}[1]{{\label{tab:#1}}}

\newcommand{\Tabff}[1]{{\ref{tab:#1}}}
\newcommand{\Tab}[1]{{Table~\Tabff{#1}}}

\newcommand{\pan}[1]{{\textrm{#1}}}

\newcommand{\FIGFF}[2]{{\ref{fig:#2}\pan{#1}}}
\newcommand{\Figff}[1]{{\FIGFF{}{#1}}}
\newcommand{\FIG}[2]{{Fig.~\FIGFF{#1}{#2}}}
\newcommand{\Fig}[1]{{\FIG{}{#1}}}
\newcommand{\FIGS}[2]{{Figs.~\FIGFF{#1}{#2}}}
\newcommand{\Figs}[1]{{\FIGS{}{#1}}}
\newcommand{\Sectff}[1]{{\ref{sec:#1}}}
\newcommand{\Sect}[1]{{Sect.~\Sectff{#1}}}

\begin{document}
\begin{center}
Submitted to {\em The Astrophysical Journal}
\end{center}
\title{Supernovae, Jets, and Collapsars \\ }

\author{A. I. MacFadyen, S. E. Woosley, and A. Heger\\ }

\vskip 0.2 in
\affil{Department of Astronomy and Astrophysics \\
University of California, Santa Cruz, CA 95064 \\ }
\authoremail{andrew@ucolick.org, woosley@ucolick.org, alex@ucolick.org}

\begin{abstract} 

We consider the explosion of supernovae and the possible production of
a variety of high energy transients by delayed black hole formation in
massive stars endowed with rotation. Following the launch of a
``successful'' shock by the usual neutrino powered mechanism, the
inner layers of the star move outwards, but lack adequate energy to
eject all the matter exterior to the neutron star. Over a period of
minutes to hours a variable amount of mass, $\sim$ 0.1 to 5 M\sun,
falls back into the collapsed remnant, often turning it into a black
hole and establishing an accretion disk. The accretion rate,
$\sim$0.001 to 0.01 M\sun \ s$^{-1}$, is inadequate to produce a jet
mediated by neutrino annihilation, but similar to that invoked in
magnetohydrodynamic (MHD) models for gamma-ray bursts (GRBs). We thus
consider the effect of jets formed by ``fallback'' in stars that are
already in the process of exploding. We justify a parameterization of
the jet power as a constant times the mass accretion rate, $\epsilon
\dot {\rm M} c^2$, and explore the consequences of $\epsilon$ = 0.001
and 0.01. Adopting an initial collimation half-angle of 10 degrees, we
find that the opening of the jet as it propagates through the
exploding star is strongly influenced by the jet's initial pressure
and the stellar structure. Cold jets tend to stay collimated, and
become even more so, sometimes having an angle of only a few degrees
when they reach the surface.  Jets having higher internal pressure
than the stellar material through which they pass, or less initial
collimation by the disk, spread out and tend to make energetic,
asymmetric supernovae accompanied, in helium stars, by weak GRBs.  SN
1998bw may have been such an event. In supergiants, shock breakout also
produces bright x-ray transients that might be a diagnostic of the
model, but even the most powerful jets (equivalent isotropic energy
10$^{54}$ erg) will not produce a GRB in a red supergiant. For such
Type II supernovae we find a limiting Lorentz factor of $\Gamma \approx
2$.  Jets produced by fallback should be more frequent than those made
by the prompt formation of a black hole and may power the most common
form of gamma-ray transient in the universe, although not the most
common form seen so far by BATSE.  Those are still attributed to
prompt black hole formation, but it may be that the diverse energies
observed for GRBs so far reflect chiefly the variable collimation of
the jet inside the star and a consequently highly variable fraction of
relativistic ejecta. Indeed, these events may all have a common total
energy near 10$^{52}$ erg.

\end{abstract}


\vfill\eject

\section{Introduction}
\lSect{intro}

At the heart of many current models for GRBs, one finds a black hole
accreting rapidly through a disk (e.g., Thompson 1994; Katz 1997;
Piran 1999; M'esz'aros 1999; MacFadyen \& Woosley 1999, henceforth,
MW99; Fryer, Woosley, \& Hartmann 1999). Models may differ in the way
the hole comes to exist, in the expected accretion rate, and,
especially, in the physical process(es) invoked to convert some
fraction of the disk's binding energy into relativistic outflow.  All
agree, however, that some non-trivial fraction, $\epsilon_1$,
($\ltaprx 6-42$~\%, depending upon the Kerr parameter) of the
accreted mass-energy $\dot {\rm M} c^2$, with $\dot {\rm M}$ the
accretion rate through the disk, goes into powering jets. In addition,
some fraction, $\epsilon_2$, of the rotational energy of the black
hole ($\ltaprx$9\% M$_{\rm BH} c^2$) may be extracted in a usable form
(Blandford \& Znajek 1977; MacDonald et al. 1986; M`esz`aros \& Rees
1997; M`esz`aros 1999; Lee, Wijers, \& Brown 1999). If these sorts of
models are to explain GRBs with equivalent isotropic energies of up to
10$^{54}$ erg and $\sim$1\% beaming factors using stellar mass black
holes and accretion reservoirs, the total efficiency, $\epsilon =
\epsilon_1 + \epsilon_2$, for GRBs cannot be much less than 1\%. This
is the limit realized in the most efficient neutrino powered models
(Popham, Woosley, \& Fryer 1999; Janka et al. 1999ab) and is
conservative compared to what is often invoked for MHD powered bursts
from merging neutron stars.

In a previous paper (MW99), we explored the collapsar model as
a possible explanation for common GRBs.  The key element in this model
was the ``failed'' explosion of a massive star whose iron core had
collapsed, first to a neutron star, then, following rapid accretion, to
a black hole (see also Woosley 1993). The black hole formed within a
few seconds of core collapse, about the same time as infalling matter
in the equatorial plane was slowed by rotation and piled into a
disk. No outward moving shock was generated prior to this time. We
shall refer to this as the ``prompt black hole'' version of the
collapsar model, or a ``Type I collapsar''.

It is also quite possible to form a black hole over a longer period of
time in a supernova that launches an outgoing shock with inadequate
strength to eject all the helium and heavy elements outside the
neutron star (Woosley \& Weaver 1995; Fryer 1999). Fryer estimates
that this sort of behavior might start around $\sim 20$ M\sun \ and
persist up to 40 M\sun. Depending upon the initial mass of the neutron
star and the equation of state, the reimplosion of from $\sim0.1$ to
several M\sun \ of the stellar mantle will be adequate to produce a
black hole. Excess matter will then accrete into the hole, and, if it
has sufficient angular momentum, will form a disk.  We shall refer to
this as the ``delayed black hole'' version of the collapsar model or a
``Type II collapsar''. Type II is probably a more frequent event than
Type I because it involves a more densely populated portion of the
stellar mass function.

As we shall see (\Sect{init}), the accretion rate when most of the
matter falls back in the delayed black hole case is typically one to
two orders of magnitude less than in the prompt case, i.e.,
$\sim$0.001 to 0.01 M\sun \ s$^{-1}$. This is too low for neutrinos to
effectively extract disk energy and form a jet (Popham, Woosley, \&
Fryer 1999; MW99). However, it is quite similar to the accretion rate
often invoked for MHD models of GRBs, especially in the case of low
disk viscosity (e.g., M`esz`aros 1999), when $\sim$0.1 M\sun \
accretes in 10 s. This motivates us to consider the possibility that
energetic jets will still form in a supernova that is already in the
process of exploding. The accretion time scale is much longer than in
the earlier collapsar model, but the total accreted mass and its
angular momentum can be comparable. If the star is a red supergiant,
the jet will overtake the supernova shock before it reaches the
surface, but since the velocity of the jet head is sub-relativistic,
the jet may lose its energy input at its base before breaking free of
the star. Little highly relativistic matter will be ejected. For a
helium star, however, the jet breaks out while continuing to receive
accretion power at its base. Its motion may become highly
relativistic.  This may also be possible for blue super giants
(R$\approx$ 50 R\sun).

The energy of the jet and the explosion it produces depend upon
the efficiency of MHD processes in extracting accretion energy from
the disk. As noted above, this is uncertain, but might be as large as
$\sim$10\%. In this paper, we shall make the simple {\sl ansatz} that
the jet power, at any point in time, is an efficiency factor,
$\epsilon$, times $\dot {\rm M} c^2$, with $\epsilon \sim 0.001 -
0.01$ (see \Sect{jete}). Then the energy potentially available for
making a jet when only one solar mass is accreted is $\sim 10^{51} -
10^{52}$ erg. This conservative estimate is still large compared to
the energy of a typical supernova and, if collimated into a small
fraction of the sky, is also enough energy for a very powerful GRB.

We thus explore in \Sect{numerical} the explosion of massive stars
endowed with rotation whose iron core collapse launches a weak shock
and, after some delay, a powerful jet. The jet has a power given by
the accretion rate and an efficiency factor, though the jet may also
react on the star to inhibit its own accretion. In addition to its
power, the jet is also defined by the radius at which it is initiated,
its opening angle, and its internal pressure. The pressure turns out
to be an important parameter as it affects the collimation properties
of the jet.  If the jet pressure is large compared to the stellar
surroundings in which it propagates, the jet will diverge. If it is
less, the jet may, under some circumstances, be hydrodynamically
focused to a still smaller opening angle.

Depending upon the collimation properties of the jet, its duration,
and total energy, one expects a wide variety of phenomena ranging from
GRBs of diverse energy and duration to energetic, asymmetric
supernovae, all powered by the same basic mechanism, a hyperaccreting
black hole (\Sect{conc}).  In all cases, the jet explodes the star in
which it propagates, but the energy imparted to the star by the jet, as
well as the maximum ``equivalent isotropic energy'' of the jet along
the axis, are highly variable, even for jets that have the same total
energy.

\section{Initial Models and Fallback}
\lSect{init}

As has been known for some time (Colgate 1971, Woosley 1988), it is
quite possible to produce a fairly massive black hole in an otherwise
successful supernova explosion - one that would be about as bright,
optically, as one that left a neutron star. Initially, the collapse of
the iron core produces a neutron star and launches a shock, but,
especially for more massive helium cores, the shock lacks adequate
energy to eject all the matter outside the neutron star. The
gravitational binding energy of the helium core increases with mass
roughly quadratically while the explosion energy does not (Fryer
1999). Some portion of the matter outside the neutron star initially
moves out, then falls back.

The rate of this fallback is particularly high when the shock wave
plows into regions with increasing $\rho r^3$ (Bethe 1990; Woosley \&
Weaver 1995). So long as the shock remains in sonic communication with
the origin, its deceleration in these regions is communicated back to
the inner mantle, a portion of which fails to achieve escape
velocity. A quantitative description of fallback must also include
the {\sl internal} energy of the ejected zones. At early times,
especially a few seconds, the pressure support provided, e.g., by a
piston, plays an important role. Internal energy is converted into
expansion energy and zones, pushing on the piston, escape that, based
solely upon their kinetic energy budget and gravitational potential,
would not have. Eventually, however, the expansion becomes supersonic
as the sound speed in the ejecta declines. Matter loses sonic
communication with the piston and evolves independently. Careful
treatments of both shock propagation and the inner boundary condition
are thus essential in any study of fallback.

Heger, Woosley, \& Langer (1999) have recently calculated the
evolution and simulated the explosion of a grid of $15$ and
$25\,\Msun$ main sequence stars, including the effects of rotation and
mass loss, for several assumptions regarding efficiencies of
semi-convection and rotationally induced mixing.  We examine here two
of their $25\,\Msun$ models.  One, with relatively inefficient
semi-convective mixing, ended its life with a low density hydrogen
envelope of $6.57\,\Msun$ (Model A).  The other, with more efficient
mixing, had an envelope of only $1.69\,\Msun$ (Model B). The helium
core mass in each was $8.06\,\Msun$ (A) and $8.87\,\Msun$ (B); the
iron core mass, 1.90 M\sun \ (A) and 1.81 M\sun \ (B); and the stellar
radius, $8.15 \times 10^{13}$ cm (A) and $8.65 \times 10^{13}$ cm
(B). The presupernova structure, composition, and angular momentum
distribution of Model A, which will be used for most of this paper,
are shown in \Fig{presn}. Note that Heger, Woosley, \& Langer (1999)
assumed rigid rotation on spherical shells, hence the actual angular
momentum in the equatorial plane is 50\% higher than in \Fig{presn})).
Both models had sufficient angular momentum to form a Kerr black hole
in the center and for the matter in the equatorial plane to form an
accretion disk around it.  That is, there is sufficient angular
momentum available for the black hole to stay at maximum rotation, and
the angular momentum in the equatorial plane is high enough to stop
infall at the last stable orbit (in co-rotation) around a Kerr black
hole.

For this paper, we calculated a new series of supernova explosions
based upon the models of Heger et al., but with a range of kinetic
energies at infinity (\Tab{Expl}). The explosion was simulated by a
piston that first moved inwards for $0.45$ s until it reached a radius
of 500 km and then moved outwards with an initially highly supersonic
velocity. This velocity was decelerated like a test particle in the
gravitational field of central mass of $\aP$ times that of the
core. For Model A the assumed core below the piston was 1.97 M\sun \
and for Model B, 1.84 M\sun. The initial velocity was chosen so that
the piston came to a rest at $10,000\,\km$ (Woosley \& Weaver 1995;
Heger, Woosley, \& Langer 1999).  An exception was Model A17 in which
a terminal radius of 100,000 km was assumed for comparison.

None of these were ``failed supernovae'' in the sense described by
Woosley (1993) and MW99.  While neutrino transport was not modeled,
the motion of the piston gave an outgoing shock that, even for the the
lowest energy considered, ejected at least the hydrogen envelope of
the star with supernova-like speeds.  Each of the models in \Tab{Expl}
would have had made a bright Type II supernova without any additional
energy input, though many would have lacked the characteristic
``radioactive tail'' from $^{56}$Co decay.  For piston trajectories
that give kinetic energies at infinity above about $1.2\Foe$,
calculations by Blinnikov (1999) have shown that the high mass
envelope star (Model A) would produce a rather typical Type IIp
supernova. The low mass envelope gives a curve more like a Type II-L
supernova (Model B).

However, especially for energies below $\sim1.0 \times 10^{51}$ erg,
large amounts of material failed to escape and fell back to the center
of the explosion.  fallback masses ranged from approximately zero to
essentially the entire helium core (\Tab{Expl}). The accretion was
followed in some detail using two quite different codes: one, the same
one-dimensional Lagrangian hydrodynamics code, KEPLER (Weaver,
Zimmerman, \& Woosley 1978), used to simulate the presupernova
evolution and the explosion. The other was a one-dimensional version
of the Eulerian code, PROMETHEUS (Fryxell, M\"uller, \& Arnett 1989,
1991; MW99). PROMETHEUS is an implementation of the high order Godunov
hydrodynamics scheme, PPM ({\bf p}iece-wise {\bf p}arabolic {\bf
m}ethod), and we shall henceforth refer to PROMETHEUS as ``the PPM
code''.

The effect of various inner boundary conditions on the fallback
was then studied. In KEPLER, the inner boundary was the piston zone, held
fixed at its final radius, usually 10,000 km. Material that fell back,
accumulated and came to rest. It was distinguished from outgoing
material by having velocity less than 100 km s$^{-1}$ and a large
positive velocity in the next zone out.

When the same calculations were repeated using PPM, the results
depended sensitively upon the treatment of the inner boundary condition
and the time the calculation was linked from KEPLER. For a
transmitting boundary condition (zero radial gradient of all
variables), initiated as the shock left the carbon-oxygen core (10 s
after core bounce), a rarefaction overtook the outgoing matter and
more matter fell back than in the equivalent KEPLER calculation. For
example, fallback in Model A11 was increased from 0.47 M\sun \ in the
KEPLER calculation to 1.0 M\sun \ in the PPM calculation, and the
explosion energy at 10,000 seconds was reduced from 1.21 $\times
10^{51}$ ergs to 1.08 $\times 10^{51}$ ergs.  

It is not physical for the matter to rest on the piston, nor is it
physical to remove all pressure support from the model while the
explosion is still developing.  We thus sought a
compromise. \Fig{wilson} shows the radial location of Lagrangian zones
as a function of time for the KEPLER (piston-driven) explosion of
Model A01. \Fig{rhor3} provides an important detail, regions in the
presupernova star where an outgoing shock wave is expected to speed up
or slow down. While initially (t~$\ltaprx$ 60 s) all zones are moving
rapidly outwards, by about 100 s a bifurcation appears and, by 250 s,
2.6 M\sun has fallen back onto the piston. This material would
collapse to the origin were the piston removed.  We thus chose to make
our link for Model A01 to PPM at 100 s, using a transmitting boundary
condition thereafter. This resulted in good agreement between the
total fallback rate calculated with the two codes (\Tab{Expl}), while
the PPM calculation more physically described the fallback of matter
to the origin at late times.

In addition to the need for hydrodynamical matching between the two
codes, the mapping at 100 s also has a physical basis.  During
approximately the first 10 s, a neutrino driven wind powered by the
radiating proto-neutron star pushes against the outgoing stellar
mantle and prevents its reimplosion.  Eventually, however, post-shock
gas which has failed to achieve escape velocity is pushed back toward
the proto-neutron star, both by gravity and an inwardly-directed
pressure gradient. This falling material overcomes the resistance of
the low density neutrino heated bubble and penetrates, perhaps by
forming ``fingers'', back to the hard surface of the neutron star
where it accumulates in an atmosphere on top of the hard neutron star
surface (\cite{fry96}, Chevalier 1989).  For the accretion rates
relevant for the first $\approx$ 100 s (\Fig{mdot}), an ``atmosphere''
builds up out to several hundred kilometers, at which point neutrino
cooling behind the shock balances the accretion energy deposition rate
and accreting gas settles subsonically onto the neutron star.  By 100
s after core collapse, enough gas has fallen back for weak explosion
energies (especially for Models A01 - A03) that the neutron star
collapses to a black hole.  For intermediate explosion energies the
central object may still be a neutron star, but the accretion rate is
such that the accretion shock forms interior to our inner boundary at
10$^9$ cm so that gas near the boundary is in free fall and the
neutral inner boundary condition is justified.  Only for accretion
rates below a few times 10$^{-6}$ M\sun \ s$^{-1}$, not considered
here, would the accretion shock move out beyond 10$^9$ cm and the
boundary condition would have to be modified.

Once a black hole has formed, gas that falls back with sufficient
angular momentum will settle into a disk.  While this part of the star
is not modeled in the current calculations, Popham et. al. (1999) and
MW99 found that the inner disk can transport the mass it receives from
the collapsing star in a steady state.  An accretion shock forms near
the Keplerian support radius of the accreting gas, typically several
hundred kilometers.  The upstream collapsing star is unaware of the
accretion disk and collapses in free fall exterior to this region.
Once again an absorbing boundary condition at 10,000 km is justified.

The accretion rates obtained from the PPM calculations are plotted as
a function of time in \Fig{mdot}. The characteristic time for half the
mass to fallback is given in \Tab{Expl}.  Accretion rates between
10$^{-4}$ and 10$^{-2}$ M\sun \ s$^{-1}$ are maintained for hundreds to
thousands of seconds.

Accretion onto collapsed objects inside of supernovae has been
considered previously in a semi-analytic fashion by Chevalier (1989)
who pointed out, for the high accretion rates considered here, that
accretion onto a neutron star would be mediated by neutrino losses
(see also Fryer, Benz, \& Herant (1996) and Zampieri et
al. 1998). Unless the neutron star magnetic field is unusually strong,
the accretion rates we are interested in are sufficiently high
that the neutron star magnetic field would be crushed to the surface
and sufficiently low so as not to launch a second supernova shock by
neutrino absorption. Chevalier gives an approximate formula for the
fallback accretion rate based on the parameters of SN~1987A and
assuming a time sufficiently late that the ``reverse shock'' has
already reached the center:
\begin{displaymath}
  \dot{\mathrm{M}} = 1.0\times\Ep{35} \, t^{-5/3} \;\g \, s^{-1}
\;.
\end{displaymath} 

For models with large amounts of fallback, our calculations show that
accretion occurs without the mediation of a strong ``reverse
shock''. Indeed, it is not possible for the reverse shock in a red
supergiant to make it back to the center of the star in only a few
hundred seconds. Consequently, Chevalier's formula becomes applicable
much earlier than the $\Ep4$ s he estimated for SN~1987A.  However,
different regions of the star have variable entropy, in particular
jumps at the boundaries of active burning shells and convective
shells. One large jump is located at the oxygen burning shell in the
presupernova star.  As a result, one sees in {\Fig{mdot}}, an early
high accretion rate that slows abruptly after the silicon shell has
fallen in (mass coordinate $2.6\,\Msun$ in the presupernova star).
After that point though, during the accretion of the carbon-oxygen
core, $\sim 400\,\Sec$ in Model A04, the accretion rate does start to
decline roughly as $t^{-5/3}$ and is in rough quantitative
agreement with Chevalier's estimate for SN~1987A.

\section {Jet Energy}
\lSect{jete}

From many observations of all sorts of situations where material
accretes from a disk into a compact object -- star forming regions,
active galactic nuclei, microquasars, SS-433, and planetary nebulae --
it seems that jets are a ubiquitous phenomenon (e.g., Livio 1999;
Pringle 1993). In each of these phenomena, the jet typically carries
away from $\sim$3\% to 30\% of the binding energy of the disk and the
jet speed is of order the escape velocity. For jets produced near the
event horizon of an accreting black hole, the relevant speed is the
speed of light. In the case of black hole accretion, one also has the
advantage of knowing the binding energy of the last stable orbit as a
function of Kerr parameter and black hole mass. The rate at which
matter flows through that last stable orbit and liberates some of
this energy in a useful form is just given by the accretion rate. This
suggests writing the power of the jet in a simple parametric form
$\dot E_{\rm jet} \ = \ \epsilon \dot {\rm M} c^2$. 

For example, for the Blandford-Znajek (1977) process for extracting
rotational energy from a black hole with Kerr parameter $a \equiv
Jc/GM^2$ predicts:
$$\dot E_{\rm jet} \ \approx \ 3 \times 10^{51} \, a^2 \, ({{\dot {\rm M}}
\over {\rm 0.1 \ M_{\scriptscriptstyle \odot} \ s^{-1}}})\: {\bf erg\,s}^{-1},$$
or in terms of our efficiency parameter, $\epsilon \approx 0.01 \, a^2$.
This assumes the development of a nearly equipartition magnetic field, 
about 10$^{15}$ gauss for stellar mass black holes. 

A similar expression that makes the dependence upon magnetic field
strength explicit is given by McDonald et al. (1986),
$$\dot E_{\rm jet} \ \approx \ 10^{50} \, a^2 \, ({{\rm M_{BH}} \over {3 \,
{\rm M_{\scriptscriptstyle \odot}}}})^2 \, ({{\rm B} \over {10^{15} 
{\rm gauss}}})^2 \: {\bf erg\,s}^{-1}.$$
For a steady state disk, $\dot {\rm M} = 4 \pi {\rm r H} \rho {\rm
v}$, with H, the disk scale height and v, the radial velocity. The
viscous time scale is approximately $\tau = r^2/\alpha H^2 \Omega_K$,
with $\alpha$, the disk viscosity parameter and $\Omega_K$, the
Keplerian angular velocity. Further, assuming a fraction, $\delta$, of
the equipartition magnetic field energy density, $B^2/4 \pi \sim \delta
\rho \Omega_k^2 r^2$, an advection dominated disk with $H \sim \beta
r$, and a disk radius where the energy is mostly generated equal to 
$\gamma$ times the Schwarzschild radius, a similar relation is again
found between the jet energy and accretion rate,
$$\dot E_{\rm jet} \ \sim \ 0.02 \ ({{a} \over {0.5}})^2 \, ({{0.5} \over
{\beta}})^3 \, ({{3} \over {\gamma}})^{2.5} \, ({{\delta} \over
{0.01}}) \, ({{0.01} \over {\alpha}}) \ \dot {\rm M} c^2.$$ Again
$\epsilon \sim 0.001$ to 0.01 is reasonable. However, this formula
makes clear the sensitive dependence of $\epsilon$ on a number of
uncertain parameters.

Similar powers can also be produced by the magnetosphere of the
accretion disk and by winds accelerated by viscous dissipation in the
disk (MW 1999; Stone, Pringle, \& Begelman 1999).  In fact, these
powers may dominate the energy extracted from black hole rotation in
most situations of interest (Blandford \& Znajek 1977; Livio, Ogilvie,
\& Pringle 1999).  There, one also expects the energy extraction to
scale with the energy density in the magnetic field (times the disk
area and Alfven speed). Using the same assumptions as the previous
paragraph, this situation also gives a jet luminosity that scales as
$\dot {\rm M} c^2$.

\section {Two Dimensional Simulation of Jet Propagation}
\lSect{numerical}

Given a prescription for the jet power and a background star in which
to propagate it, one can then model the interaction of the jet with
the star.

As a matter of convenience, we separate our discussions and
calculations into two regions, one inside 10$^9$ cm, another from
10$^9$ cm to the surface of Model A01 at $8.15 \times 10^{13}$ cm.
This segregation reflects, in part, the difficulty of carrying too
large a dynamic range on the computational grid. Time steps imposed by
the Courant condition at, e.g., $\sim$ 10$^7$ cm, are too small to
follow the progression of the explosion to the surface at $\sim$
10$^{14}$ cm. The split also separates the problem into regions of
different physics. In the inner region, the jet forms, with all the
attendant uncertain physics that involves. The outer region, however,
contains the bulk of the mass and volume of the star. Occurrences
inside 10$^7$ cm are discussed in \Sect{initjet}. The numerical models
presented there are still subject to the Courant restriction which
makes it difficult to follow fallback for hundreds, or even thousands
of seconds, so our arguments in that section are based upon the
stellar and disk structure calculated for {\sl Type I} collapsars in
MW99. Expected differences between Type I and Type II are discussed,
but we are mainly interested in this region as providing reasonable
inner boundary conditions for the outer region treated in \Sect{twod}.

In the outer region, the jet may be defined by an (angle-dependent)
flux at the inner boundary of energy and density. Studies of this
region will show how jets of different properties affect the star
through which they propagate, in particular how much energy is shared
with the star, whether the jet suffers degradation or additional
collimation, and the degree to which relativistic matter is ejected.
Note that though the results of \Sect{twod} are discussed within 
the context of the collapsar model for GRBs, they are equally
appropriate to any model that produces jets with similar properties,
for example jets powered by a highly magnetic, rapidly spinning
neutron star (Usov 1994; Wheeler et al. 1999, Khokhlov et al. 1998).

\subsection {Jet Initiation}
\lSect{initjet}

A variety of processes may operate in the vicinity of a rotating black
hole to create a jet. Very close to the hole, magnetic fields
maintained by currents in the disk and threading the black hole
ergosphere can extract rotational energy - the Blandford-Znajek (1977)
mechanism. Farther out, other MHD processes may also extract angular
momentum and energy from the disk to form a jet, e.g., by
magneto-centrifugal forces (Blandford \& Payne 1982). See Koide,
Shibata, \& Kudoh (1998), Meier (1999), Koide et al. (1999) and
Krasnopolsky, Li, \& Blandford (1999) for recent discussions. These
processes tend to produce relatively cold jets, in the sense that the
thermal energy of the jet is not initially large compared to either
the rest mass or the jet kinetic energy. There may be a critical
magnetic field above which these mechanisms are able to provide
directly matter with high Lorentz factor (Meier et al. 1997). Other
processes like neutrino energy deposition (Woosley 1993; MW99; Janka
et al. 1999ab) or magnetic reconnection (Thompson 1994; Katz 1997) in
the disk deposit their energy chiefly as heat and make hot jets. In
these cases the initial velocity of the jet is small and asymptotic
velocity is given by the initial ratio of thermal energy to rest
mass. Because the pressure from the deposited energy is initially
isotropic, the collimation of hot jets then occurs as a consequence of
the pressure and density gradients of the matter in which energy is
deposited. Cold MHD jets may thus be collimated both by magnetic
fields and by geometry, but hot jets are collimated only by the
structure of the medium in which they expand.

The jets of MW99 are hot jets, focused by the thick accretion disk and
declining radial pressure gradient in the region where energy was
deposited. In order not to obtain a velocity greatly in excess of the
speed of light in their Newtonian code, MW99 were restricted to
depositing energy at a rate that, in steady state, gave an internal
energy to baryon loading less than 10 times the rest mass (even this
gave speeds of over 10$^{11}$ cm s$^{-1}$!). These jets easily
penetrated the helium core in which they were produced while
concurrently blowing it up, but details of shock break out, and
especially a first principles determination of the asymptotic Lorentz
factor of the jet, were obscured by use of a hydrodynamics code that
was not special relativistic. More recently, Aloy et al (1999b) (see
also M\"uller et al. 1999) have recalculated the MW99 conditions using
a fully relativistic version of the PPM code (Aloy et al. 1999a) and
obtained similar results - i.e., relativistic jet penetration of the
helium core. In addition, they were able to determine that, {\sl at
breakout}, Lorentz factors of 20-30 were achieved in the jet. It is
expected that higher values may be achieved by running the calculation
longer. These hot jets were all in the context of a Type I collapsar.

It is not clear whether the jets made by fallback in a Type II
collapsar will be hot or cold (although jets mediated by neutrino
transport are impossible). Given the continuing limitations of our
Newtonian treatment, we can only explore the physics of mildly
relativistic ``warm'' jets. However, we can vary such things as the kinetic
energy and internal energy loading of the jet, its initial
collimation, and the mass and structure of the accretion disk in which
it propagates in order to understand better how the jet is dynamically
focused. 

As an example, we consider jets initiated in the two accretion disk
structures studied by MW99, a high viscosity disk ($\alpha = 0.1$,
their Fig. 8 at 7.6 s) and a low viscosity disk ($\alpha$ = $5 \times
10^{-4}$, their Fig 22 at 9.4 s). The masses of these disks differ by
approximately a factor $\alpha^{-1}$, a few hundred. One might expect
a similar difference in disk mass for constant viscosity, but reducing
the accretion rate by a few 100 (Popham et al. 1999). Thus the
difference between jet behavior initiated in these two models
qualitatively illustrates the effect we expect for the high accretion
rate in Type I collapsars and the two order of magnitude lower
accretion rate in Type II collapsars.

Jets were initiated in both models at an inner boundary radius of 50
km, with an opening half-angle of 10 degrees, a velocity of 10$^{10}$
cm s$^{-1}$, and a total power of $1.8 \times 10^{51}$ erg
s$^{-1}$. Half of the injected energy was in kinetic energy and the
other half in internal energy. The energy to rest mass ratio was thus
about 10\% so as to avoid super-luminal velocities on the
grid. \Figs{jetinit} and \Figff{flamejet} show the results after 0.6 s of
jet propagation.

Both calculations clearly show the effect of geometric focusing, but
the jet in the low mass accretion disk is much less collimated. Since
the sub-relativistic gas has comparable internal and kinetic energies,
the sound speed is near the radial streaming speed, so one expects the
jet, in a vacuum, to have comparable theta and radial velocities. This
is initially the case for the calculation using the low
disk-mass. After a thousand km, the opening half-angle is about 20
degrees. The pressure gradient of the star is still enough to maintain
some mild collimation. In the high disk-mass case though, the opening
angle is about half as great, comparable to the 10 degree input
angle. If one continued this trend to still smaller disks and higher
internal energy loading factors, the jet would fan out still
more. Obviously, in the limit of no disk and energy deposited entirely
as a thermal bomb at the middle, an isotropic explosion would
develop. An interesting question still to be explored is what would
happen in the absence of a disk, but for off-center energy deposition.

The calculations of MW99 (their Fig. 26) showed that the ratio of
internal energy to mass density in the jet declined roughly as
r$^{-1/3}$. That is, the density declined between r$^{-1}$ r$^{-2}$ and
the radiation entropy, T$^3$/$\rho$, was constant. For constant mass
flux, constant velocity, and constant opening angle, $\theta$, the
quantity $\pi r^2 \theta^2 \rho v$ should be a constant, so that $\rho
\propto r^{-2}$. That the density declined more slowly reflects the
hydrodynamical collimation and mild deceleration of the jet. Assuming
that the r$^{-1/3}$ scaling persists for much lower values of initial
internal energy, one expects the ratio of internal energy to kinetic
energy to decline by about a factor of 10 going from the region where
the jet forms out to 10$^9$ cm. So for jets that initially had from, say,
1\% to 100\% of their initial energy in the form of radiation, the
ratio of pressure to kinetic energy, $f_{\rm P}$, at 10$^9$ cm would
be in the range 0.001 - 0.1. We shall use these values in the next
section. For much hotter jets, the effects of special relativity
(especially the contribution of pressure to momentum) make the above
scaling law invalid and the jets stay much more tightly collimated
than a non-relativistic calculation would suggest (Woosley, MacFadyen,
\& Heger 1999; M\"uller et al. 1999; Aloy et al. 1999ab). For much
cooler jets we shall find that the answer does not vary much for
f$_{\rm P}$ less than 0.01 (\Fig{press}).

\subsection {Jet Propagation and Supernova Explosion}
\lSect{twod}

The spherically symmetric explosion of Model A01, followed until 100 s
after the launch of a weak shock in the KEPLER code (\Sect{init}), was
mapped onto the Eulerian grid of a two-dimensional version of the PPM
code. This grid used 200 radial zones spaced logarithmically between
an inner boundary at 10$^9$ cm and the stellar surface at
8.1$\times$10$^{13}$ cm.  Forty angular zones, concentrated near the
pole, were used to simulate one quadrant of the stellar volume,
assuming axial and reflection symmetry across the equatorial
plane. The angular resolution varied from 1.25$^\circ$ at the pole to
3.5$^\circ$ at the equator.  Nine atomic species, (C, O, Ne, Mg, Si,
Fe, He, n, p), were carried and the equation of state included
radiation and an ideal gas consisting of electrons and the nine
ions. At 100 s, the inner 1.99 M\sun \ of the star was removed and
replaced by a transmitting (zero radial gradient of all variables)
boundary condition at 10$^9$ cm.  The 1.99 M\sun \ and all
subsequently accreted matter contributed to a point mass term in the
gravitational potential which was calculated using an integral Poisson
solver (M\"uller \& Steinmetz, 1995, MW99). At this time, the weak
initial shock was already at 1.1$\times 10^{10}$ cm when a simulated
jet was turned on at the inner boundary.

A given jet powered model was specified by its energy flux, $F_{\rm e}(t)$,
mass flux, $F_\rho(t)$, momentum flux, $F_{\rm u}(t)$, and the pressure of
the jet, $P_{\rm jet}$, all at 10$^9$ cm,

\begin{eqnarray}
F_{\rm e} & = & \dot{E}_{\rm jet}(t)/A_{\rm jet} 
\\ F_{\rho} & = &2
{\rm F_e} \ \left[ f_{\rm P} \left( \frac{1}{\gamma_{\rm jet}-1} + 1
\right) + 1 \right]^{-1} v_{\rm jet}^{-2} \\ F_{\rm u} & = & F_{\rho} v_{\rm
jet} \\ P_{\rm jet} & = & \frac{1}{2} f_{\rm P} F_{\rm u}
\end{eqnarray}
with
$$ \rm \dot{E}_{\rm jet}(t) = \epsilon c^2 \times \left\{
\begin{array} {l@{\quad:\quad}l} \dot{\rm M}(t) & t < t_{\rm p} \\
\dot{\rm M}(t_{\rm p}) \rm{max}[(t/t_{\rm p})^{-5/3},10^{-6}] & t \ge
t_{\rm p}\end{array} \right. $$ Here $t_{\rm p}$ is the time when the
accretion rate begins to follow the power law decline, usually several
hundred seconds (\Fig{mdot}). The opening half-angle of the jet was
taken to be $\theta_{\rm jet} = 10^{\circ}$ and the adiabatic index,
$\gamma_{\rm jet} = \frac{4}{3}$.  The accretion rate $\dot{\rm M}$
was restricted to that matter that fell in through the inner boundary
within 45$^{\circ}$ of the equator.  Thus the density
of the jet is given by $2 \epsilon \dot{\rm M} c^2/((4 f_{\rm P} +1)
A_{\rm jet} v_{\rm jet}^3)$ where $A_{\rm jet}$ is the cross sectional
area of the initial jet, $1.9 \times 10^{17}$ cm$^2$ for the assumed
boundary radius and collimation. The assumed initial velocity of the
jet, $v_{\rm jet}$, was 10$^{10}$ cm s$^{-1}$.

We considered four cases, $\epsilon = 0.001$, $f_{\rm P}$ = 0.001,
0.01, and 0.1, and $\epsilon = 0.01$, $f_{\rm P}$ = 0.01.  The results
of these calculations are summarized in Table 2 and \Figs{ecomp} -
\Figff{equiso}. Here the model naming follows the convention ``JMN'' where
``J'' indicates the model included a jet, but was otherwise based upon
the initial Model A01 (Table 1), ``M'' is the exponent of the
efficiency factor, $\epsilon = 10^{\rm -M}$, and ``N'', the exponent
of the pressure factor, $f_{\rm P} = 10^{\rm -N}$.  The accreted mass,
$\Delta {\rm M}$ in Table 2, is smaller than the 3.71 M\sun \ computed
without a jet (\Fig{mdot} and Table 1) for Model A01, because the jet
impeded the accretion at high latitude and because the accretion was
not quite complete after 500 seconds (\Fig{mdot}). The total energy
input by the jet was still $\epsilon \Delta {\rm M} c^2$, but the
energy in Table 2 was also reduced by the work done up to 500 s in
unbinding the star and by the internal and kinetic energy which passed
inside the inner boundary.  The $2.55 \times 10^{50}$ erg due to the
initial shock has been subtracted in Table 2 so that E$_{\rm tot}$
reflects only the energy input by the jet.

In all cases a very energetic asymmetric supernova resulted. Since the
integrated mass of the jet in our code was comparable to that of the
stellar material within 10 degrees, the time for jet break out was
approximately the stellar radius divided by the jet input speed, or
about 8000 s. Since the energy of the jet engine had declined greatly
by that time, due to the declining accretion rate (\Fig{mdot}), the
jet that broke out was only mildly relativistic (see also
\Sect{transients}). Both the long time scale and the small amount of
relativistic matter are inconsistent with what is seen in common
GRBs. However, if the hydrogen envelope had been lost leaving a bare
helium star, a longer than typical GRB could have resulted.

\Figs{press} and \Figff{equiso} illustrate how the pressure balance
between the jet and the star through which it propagates affected its
collimation properties. The interaction at late times with the
hydrogen envelope has a smaller effect on the angular energy
distribution which was set chiefly by $f_{\rm P}$ and the interaction
with the helium core.  Model J33 had the lowest internal pressure
(note that the actual value of the initial pressure depends upon the
product of $\epsilon$ and $f_{\rm P}$). The final jet was collimated
even more tightly than given by its initial injection. That is, a jet
initially of 10 degrees half width will exit the star with a FWHM of
less than two degrees, about 0.06\% of the sky, though the angular
resolution of the code is questionable for such small
angles. Meanwhile the energy at larger angles was not much greater
than that given by the initial, weak spherically symmetric explosion,
10$^{50.4}$ erg. There was little sharing of the jet energy with the
star and, except for the jet, the supernova energy remained low.

This behavior is to be contrasted with Models J22 and J31 where the
jet collimation was weaker and much more energy was shared with the
star. Note that though Model J22 had about 6 times the total energy of
J31 owing to its larger $\epsilon$, the fraction of energy at large
angles in both these models was significantly greater than in Models
J32 and J33.  That Model J22 was not ten times more energetic (the
ratio of the $\epsilon$'s) shows the inhibition of the accretion by
the strong jet. Still, Model J22 would be a very powerful supernova as
well as one accompanied by a jet.

As the jet pushes through the star, a shell of relatively high density
material is built up at its head.  The shell contains material from
the inner regions of the star which is swept along as the jet
propagates through the star.  The red regions near the polar axis in
\Fig{press} correspond to the location of the shells. If some of this
material is accelerated to relativistic velocities by the jet, as
preliminary relativistic calculations indicate (Aloy et al 1999b), it
may produce observable features in the spectrum of the burst.  More
work is needed, however, to determine the detailed composition of the
jet material at large distances from the center of the star.

The angular factor R($\theta > 10^\circ$) in Table 2 is the ratio of
the integral of the kinetic energy due to the jet outside 10 degrees
polar angle (98.5\% of the sky) to the total kinetic energy in the
star due to the jet (see \Fig{equiso}).  These energies were computed
by taking the total kinetic energy at 400 s after jet initiation in
both regions and subtracting the kinetic energy of the initial
supernova shock.  R($\theta > 10^\circ$) measures the extent to which
the jet spread laterally and shared its energy with the rest of the
star.  The limiting case, R($\theta > 10^\circ$) = 0, would correspond
to a jet that shared none of its energy with the supernova outside an
initial 10$^\circ$ polar angle.  This sort of behavior is expected for
``cold'' jets with internal pressure small compared to the exploding
helium core.  The other extreme, where the jet shared its energy
evenly with the entire star and produced a spherical explosion, would
correspond to R = $\cos \theta$ = 0.985.  Our ``hot'' jets lie
somewhere between these two limits. The quantity R($\theta >
20^\circ$) was similarly computed for a polar angle of 20$^\circ$. The
isotropic limit there would be 0.940.

Most of the curves in \Fig{equiso} were evaluated at a time (500 s)
when the central part of the jet had moved well outside the helium
core (initial radius $5.0 \times 10^{10}$ cm - defined by the point
where the mass fraction of hydrogen is 1\%), but had not yet
encountered much of the hydrogen envelope mass. Thus the energy
distribution is also appropriate to exploding helium stars of mass
$\sim$9 M\sun.  These curves then give the angular energy distribution
of Type Ib and Ic supernovae that would accompany GRBs produced by
jets of the specified properties.

For one calculation, J32, however, the propagation beyond 500 s was
considered all the way to the surface of the red supergiant at 7820 s
($\sim R/v_{\rm jet}$; note the near constancy of the jet head speed
in the hydrogen envelope). This calculation required a second mapping
of the explosion onto a new grid. At 500 s a new inner boundary was
set up at 10$^{11}$ cm to alleviate the restrictive time step
limitation imposed by the Courant condition at small radii.  The mass
interior to the new inner boundary was added to the central point
mass.  150 radial zones spaced logarithmically between 10$^{11}$ cm
and 8$\times 10^{13}$ cm were used with the same angular zoning as
before.  At 500 s the bulk of the accretion had already taken place
(\Fig{totm}) and the subsequent accretion could be approximated using
the $t^{-5/3}$ scaling at late times (\Fig{mdot}), $\dot{M}(t) =
\dot{M}(500\,s) \times \max [(t/500)^{-5/3}, 10^{-4}]$.  The jet was injected
as a boundary condition at the new inner boundary, $r=10^{11}$ cm, as
before with $\theta_{jet} = 10^{\circ}$ and with $v_{jet} = 1 \times
10^{10}$ cm s$^{-1}$. $f_{\rm p}$ was reduced to 10$^{-3}$ to approximate
the conversion of thermal energy to kinetic energy which occurs as the
gas expands adiabatically between $10^9$ cm, where $f_{\rm p}$ was
previously $0.01$, and $10^{11}$ cm.

The results of this calculation at the time of jet breakout are shown
in \Fig{breakout}. Note the lateral propagation of a strong, high Mach
number shock with a large pressure and density jump. This shock wraps
around the star and, by about 10,000 s, has ejected even the material in
the equatorial plane. The jet core spreads significantly out to about
10$^{13}$ cm, well into the hydrogen envelope, but then is
recollimated by a ``cocoon'' of high pressure, shock heated
gas. Further out at $5 \times 10^{13}$ cm, it spreads once again due
to partial blockage by a plug of high density material which the jet
shoves along.

\section{Shock wave Break Out}
\lSect{transients}

A strong shock wave breaking out of a star that is experiencing mass
loss will produce transient electromagnetic emission in two ways -
first as the shock breaks through the surface and exposes hot material
(Colgate 1969, 1974), and second as the highest velocity material
encounters the circumstellar wind (Chevalier 1982, Fransson 1984,
Leising et al. 1994). While the highest energy ejecta in the present
models may only have a small solid angle (\Fig{equiso}), there still
may be a very high luminosity in both forms of transients. This
radiation would not be appreciably beamed.

To explore this possibility, we compute the effects of two strong,
spherically symmetric shocks in Model A01 using the KEPLER
code. The motion of the piston at the edge of the iron core (see
\Sect{init}) was adjusted to give both models very high kinetic energy
at infinity, $1.44 \times 10^{53}$ erg and $1.39 \times 10^{54}$
erg. While these were one-dimensional calculations, they should
simulate the conditions experienced by stellar regions within solid
angles that experience these ``equivalent isotropic energies''
(\Fig{equiso}). The advantage of the one-dimensional Lagrangian
calculations is that very fine zoning can be employed near the
surface and radiative diffusion can be included in the calculation.

The resulting light curves for the two models are given in
\Fig{breakout}. Since KEPLER has a very simple radiation transport
scheme (flux-limited radiative diffusion) and the opacity is
dominantly due to electron scattering, the effective temperature
obtained in these calculations is an underestimate of the actual
(color) temperature, T$_{\rm c}$, by a factor of two to three (Ensman \&
Burrows 1992). Thus we expect, for the higher energy model, a
transient with mean photon energy, 3 kT$_{\rm c} \sim$ 0.25 keV lasting for
about 10 s. Since these x-rays would not be beamed, the luminosity
would be the value in \Fig{breakout} times an appropriate solid angle
(about 1\% of the sky), or $\sim$10$^{47}$ erg s$^{-1}$. Less energetic, but
longer lasting transients would come from regions of the surface that
experienced a smaller equivalent isotropic energy. This is much
brighter and harder than the shock breakout transients expected from
common Type II supernovae and would be a diagnostic of our model.
   
Even more energy may come out, but over a longer time, in the form of
x-ray and radio afterglows from circumstellar interaction. KEPLER is a
non-relativistic hydrodynamics code, but we can nevertheless estimate
the amount of relativistic ejecta following Woosley, Eastman, \&
Schmidt (1999). Assuming the near constancy of the quantity $\Gamma
\beta (\rho r^3)^{1/5}$ (with $\beta$ = v/c and $\Gamma =
(1-\beta^2)^{-{1/2}}$) across the mildly and strongly relativistic
domains (Gnatyk 1985; McKee \& Colgate 1973), we obtain $\Gamma$ as a
function of the external mass using the output of KEPLER in the
non-relativistic region (\Fig{gamma}). The finest zones in the KEPLER
calculation had mass 10$^{29}$ g. However, external to about 10$^{31}$
g the density was influenced by a transition from optically thick to
optically thin zones, and outside of 10$^{30}$ g, the density was
affected by a 3 dyne cm$^{-2}$ surface boundary pressure employed to
stabilize the calculation. We thus only trust the calculated
distribution of $\rho r^3$ out to 10$^{31}$ g and expect that there
may be relatively minor deviations from that power law outside. Of
course, we also cannot employ zones in our fit which in the KEPLER
calculation were moving at super-luminal speeds after the conversion
of all internal energy to kinetic energy.

For these reasons, we used only the KEPLER results inside 10$^{31}$ g
and used a power law extrapolation for smaller masses. The fits for
$\Gamma \beta (\rho r^3)^{1/5}$ gave nearly constant value of $7.3
\times 10^5$ and $3.3 \times 10^6$ (units g$^{1/5}$) for the $1.44
\times 10^{53}$ and $1.39 \times 10^{54}$ erg explosions
respectively. From these we infer the distribution of $\Gamma$ with
exterior mass given in \Fig{gamma} for the high energy explosion. Much
less relativistic matter was ejected in the lower energy explosion.

One cannot use the scaling relation for $\Gamma$ indefinitely though.
Especially for extended objects like red supergiants, there comes a
point when the shock thickness, mediated by radiation and electron
scattering, becomes greater than the density scale height. At about
the same point, material behind the shock will become optically thin
and lose energy. For relativistic shocks, both of these conditions
occur when the Thompson depth of material just behind the shock
becomes unity (Imshennik \& Nadyozhin 1989). From the known dependence
of $\rho r^3$ on external mass (\Fig{gamma}), $\rho r^3 = 10^{5.31}
(\Delta {\rm M} \, (\rm g))^{\delta}$ with $\delta = 0.86$, one
determines the external mass beyond which the shock speed cannot be
extrapolated
$$\Delta {\rm M} \ = \ 2.0 \times 10^{29} \left({{R_*} \over {10^{14} \ {\rm
cm}}}\right)^2 \left({{\delta} \over {0.86}}\right) \left({{0.34 \
{{\rm cm}^2 \ {\rm g}^{-1}}} \over {\kappa}}\right) \ {\rm g},$$
where $R_*$ is the stellar radius. 
For the much more compact stars considered by Woosley, Eastman, \&
Schmidt (1999), small amounts of material can be accelerated to
comparatively large values of $\Gamma$, but here, for red supergiants,
there is a limit of $\Gamma \approx 2.5$. Thus jet powered Type II
supernovae cannot make GRBs of the common type. The duration of the
jet is too short for the jet itself to acquire large $\Gamma$ outside
the star and the relativistic shock acceleration mechanism fails to
yield sufficiently energetic material.

However, 10$^{-4}$ M\sun \ of material with $\Gamma$ = 2.5 is still
about 10$^{50}$ erg of kinetic energy. This material will radiate its
kinetic energy after encountering roughly its rest mass, which will
take a time
$$\tau \ = \ 10^4 \ \left({{10^{-5} \ {\rm M\sune \ y^{-1}}} \over
{\dot {\rm M}}}\right) \left({{10 \ {\rm km \ s^{-1}}} \over {v_{\rm
wind}}}\right) \ {\rm s}.$$ 
For a solid angle 1\% of the sky, this again implies $\sim 10^{44}$
erg s$^{-1}$, presumably in hard x-rays. In fact, the circumstellar
interaction of these breakout shocks resembles in many ways the
multi-wavelength afterglow of GRBs, except that there would be no
GRB. Since these could be frequent events and may not be beamed to the
same small angle as a GRB, there could be many such ``orphan
afterglows''.

\section{Conclusions}
\lSect{conc}

It is widely believed that many diverse phenomena in galactic
nuclei -- quasars, BL-Lac galaxies, Seyferts, radio galaxies, blazars,
etc. -- are powered by a common engine, accretion at a rate of a few
solar masses per year into a supermassive black hole of several
billion M\sun \ (Blandford \& Rees 1974; Antonucci 1993). The
differences relate to the mass and rotation rate of the hole, the
environment in which it accretes, and the angle at which one
observes. We suggest here that a similarly large variety of energetic
phenomena can also be powered by accretion at a much higher rate
into a small, stellar mass black hole. It could be that GRBs are but
the ``tip of the iceberg'' and that many other interesting kinds of
events await discovery.

In this paper, we have drawn attention to the possibility of two kinds
of hyper-accreting black hole scenarios for making supernovae and high
energy transients. Type I collapsars, as explored previously by MW99,
have black holes that form promptly. The black holes in Type II
collapsars, on the other hand, form after some delay owing to the fall
back from a supernova lacking adequate energy to eject all of its
helium core and heavy elements.

We have shown (\Fig{equiso}) that explosions of nearly constant total
energy can have highly variable ``equivalent isotropic energy''
dependent not only upon the angle at which they are viewed, but also
upon how passage through the star acts to collimate (or de-collimate)
the jet. This should be true of both Type I and Type II
collapsars. For the roughly 10 GRBs for which redshifts and GRB
energies had been determined as of July, 1999, the inferred energy has
an enormous range - from about $8 \times 10^{47}$ erg for GRB 980425
(Galama et al. 1998) to several times 10$^{54}$ erg for GRB 990123
(Kulkarni et al. 1999). Yet, as MW99 pointed out, the total explosion
energy for both these events may have been $\sim$10$^{52}$ erg with
the energy collimated into a tight jet for GRB 990123 (see also
\Fig{equiso} and Table 2) and dissipated in producing a powerful
supernova for GRB 980425 (Iwamoto et al. 1998; Woosley, Eastman, \&
Schmidt 1999). The collimation properties of the jet may be more
important than its total energy in determining the observed properties
of a GRB.

We continue to support the view of MW99 that common, long hard GRBs
(mean duration 20 s) seen by BATSE are produced in Type I
collapsars. The GRB commences only after the jet has drilled a hole
through the star (5 - 10 s) so that it can expand freely. The jet may
be energized either by neutrino energy transport from the disk or by
MHD processes.  However, we also see the possibility of a number of
other phenomena in Type II collapsars depending upon the
collimation of the jet, the duration and mass of the accretion, and
the nature of the presupernova star.

\vskip 0.2 in
\noindent
{\sl``Smothered'' and broadly beamed gamma-ray bursts; GRB 980425} -
These occur in helium stars in which the jet either fails to maintain
sufficient collimation (e.g., is too ``hot'' compared to the star
through which it propagates), or loses its energy input before
breaking out of the star ($\ltaprx$10 s; MW99, MacFadyen \& Woosley
1998). An energetic supernova still occurs (SN 1998bw, in this case)
and a weak GRB is produced, not by the jet itself, but by a strong,
mildly relativistic shock from break out ($\Gamma \sim 5$)interacting
with the stellar wind (Woosley, Eastman, \& Schmidt 1999). Kulkarni
et al. (1998) and Wieringa, Kulkarni, \& Frail (1999) have also argued
that the radio emission from GRB 980425 was not strongly beamed and
had a total energy not too much greater than the GRB ($\ltaprx
10^{49}$ erg) suggesting that GRB 980425 was {\sl not} a much more
powerful GRB observed from the side (as in e.g., Nakamura 1999).
Because these events are so low in gamma-ray energy, many could go
undetected by BATSE and these could be the most common form of GRB in
the universe. Because the initial jet may be less effectively
collimated in GRBs made by supernova fallback, it is tempting to
associate these phenomena with delayed black hole formation and the
stronger GRBs with prompt black hole formation. More study is needed,
but we tentatively offer Model J22 in \Fig{equiso} as a possible
prototype for SN 1998bw. This model at 500 s postbounce would appear
essentially as a $\sim 10^{52}$ erg He core explosion with a small
(10$^\circ$) hole along each axis. A ``hotter'' jet, lower disk mass,
or larger efficiency factor, $\epsilon$, might provide even better
agreement.

\vskip 0.2 in
\noindent
{\sl Long gamma-ray bursts; $\tau_{\rm burst} \gtaprx 100$ s} - Though
typical ``long, complex bursts'' observed by BATSE last about 20
seconds, there are occasionally much longer bursts. For example,
GRB950509, GRB960621, GRB961029, GRB971207, and GRB980703 all lasted
over 300 s. These long durations may simply reflect the light crossing
time of the region where the jet dissipates its energy (modulo
$\Gamma^{-2}$), especially in the ``exterior shock model'' for
GRBs. However, if the event is due to internal shocks, the duration
depends on the time the engine operates. Such long bursts would imply
enduring accretion on a much longer time scale than one expects in the
Type I collapsar model where the black hole forms promptly. The
fallback powered models discussed in this paper could maintain a GRB
for these long time scales (\Fig{totm}). Indeed considerable power may
still be developed by the GRB days after the initial event with a
luminosity declining roughly as t$^{-5/3}$ (\Fig{mdot}).

\vskip 0.2 in
\noindent
{\sl Very energetic supernovae - SN 1997cy} - Germany et al. (1999)
have called attention to this extremely bright supernova with an
unusual spectrum. The supernova was Type IIn and its late-time light
curve, which approximately followed the decay rate of $^{56}$Co, would
require $\gtaprx$2 M\sun \ of $^{56}$Ni to explain its
brightness. Perhaps this was a pair-instability supernova (Woosley \&
Weaver 1982; Heger, Woosley, \& Waters 1999). On the other hand,
circumstellar interaction could be the source of the energy and the
agreement with $\tau_{1/2}$($^{56}$Co) merely fortuitous. This would
require both a very high explosion energy and a lot of mass loss just
prior to the supernova. The sort of model, described in \Sect{twod},
especially Model J22, could provide the large energy in a massive star
that would be naturally losing mass at a high rate when it died. Large
quantities of $^{56}$Ni might also be made by the wind from the
accretion disk (see below). But the presupernova radius of this event
was too large and the jet would have shared its energy with too great
a mass to make a common GRB. In \Sect{transients}, we placed a limit
of $\Gamma \ltaprx 2.5$ on any relativistic ejecta in such events. We
thus regard the detection of a short, hard GRB from the location of SN
1997cy as spurious.

\vskip 0.2 in
\noindent
{\sl Nucleosynthesis - $^{56}$Ni and the $r$-process} - An explosion
of 10$^{52}$ erg focused into 1\% of the star (or 10$^{53}$ erg into
10\%) will have approximately the same shock temperature as a function
of radius as an isotropic explosion of 10$^{54}$ erg. From the simple
expression ${{4} \over {3}} \pi r^3 a T^4 \sim 10^{54}$ erg (Woosley
\& Weaver 1995), we estimate that a shock temperature in excess of 5
billion K will be reached for radii inside $4 \times 10^9$ cm. The
mass inside that radius external to the black hole (assumed mass
initially 2 M\sun) depends on how much expansion (or collapse) the
star has already experienced when the jet arrives. Provided the star
has not expanded much before the jet arrives, an approximate number
comes from the presupernova model, 3 M\sun \ times the solid angle of
the explosion divided by $4\pi$, or $\sim$0.1 M\sun.  Additional
$^{56}$Ni will be synthesized by the wind blowing off the accretion
disk (MW99; Stone, Pringle, \& Begelman 1999) and this may be the
dominant source in supernovae like SN 1998bw. Even for the relatively
low accretion rates in Type II collapsars, the temperature in the
accretion disk is hot enough to photodisintegrate the accreting matter
to nucleons (Popham et al. 1999). As this material expands and cools,
it will reassemble mostly to $^{56}$Ni. Stone et al. estimate that an
appreciable fraction of the accreted matter will be ejected in such a
wind.

The composition of the jet itself depends upon details of its
acceleration that are hard to calculate. However it should originate
from a region of high density and temperature (Popham et
al. 1999). The high density will promote electron capture and lower
$Y_{\rm e}$. The high entropy, low $Y_e$, and
rapid expansion rate are what is needed for the $r$-process (Hoffman,
Woosley, \& Qian 1997). The mass of the jet, $\sim10^{-4}$ M\sun \
(corrected for relativity) is enough to contribute significantly to
the $r$-process in the Galaxy even if the event rate was $\ltaprx$1\%
that of supernovae and the jet carried only a fraction of its mass as
$r$-process.

\vskip 0.2 in
\noindent
{\sl Soft and hard x-ray transients from shock breakout} - Focusing a
jet of order 10$^{52}$\,ergs into 1 - 10\% of the solid angle of a
supernova results in a shock wave of extraordinary energy
(\Fig{equiso}).  As it nears the surface of the star, this shock is
further accelerated by the declining density gradient. We have
estimated (\Sect{transients}, \Fig{softx}) that as the shock erupts
from the surface of the star one may have soft x-ray transients of
luminosity up to $\sim10^{49}$ erg s$^{-1}$ times the fraction of the
sky to which high energy material is ejected (typically 0.01). The
color temperature at peak would be approximately $2 \times 10^6$ K
(see also Matzner \& McKee 1999). A 10$^{53}$ erg shock gave a
transient about half as hot and ten times longer and fainter. The
impact of the mildly relativistic matter could give an enduring x-ray
transient like the afterglows associated with some GRBs, even though
the time scale is too long for the x-ray burst to be a common GRB
itself.

\vskip 0.2 in
\noindent
{\sl Supernova remnants with unusual symmetry} - The expanding remnant
of a supernova exploded by a jet should lack spherical symmetry. Along
the rotational axis of the presupernova star matter expands with much
greater velocity than along the equator. Dependent upon details of the
circumstellar shock interaction, this could either lead to an
elongated nebula characteristic of bipolar outflow or a torus of
slower moving, denser debris in the equator. The massive stars that
produce collapsars have a very high abundance of oxygen, so one might
seek either toroidal or bipolar SNRs that are oxygen-rich. Lasker
(1980) pointed out that N132D in the LMC might be such a remnant. More
recent analysis (Sutherland \& Dopita 1995; Morse, Winkler, \&
Kirshner 1995) suggests that the remnant may have elliptical structure.
Another SNR with toroidal structure is E0102-72 in the SMC recently
studied by the Chandra X-Ray Astronomy Facility.

\vskip 0.2 in
\noindent
{\sl Mixing in supernovae - SN 1987A} - It is generally agreed (Arnett
et al. 1989) that the explosion that gave rise to SN 1987A initially
produced a neutron star of approximately 1.4 M\sun.  There may have
been $\sim$0.1 M\sun \ of fallback onto that neutron star (Woosley
1988) and a black hole may or may not have formed. Again invoking our
{\sl ansatz} that ${\rm L_{jet}} = \epsilon {\rm \dot M} c^2$, even
for $\epsilon \sim$ 0.003, we have a total jet energy of $6 \times
10^{50}$ erg. This is about half of the total kinetic energy inferred
for SN 1987A. Thus very appreciable mixing and asymmetry would be
introduced by such a jet - {\sl provided the material that fell back
had sufficient angular momentum to accumulate in a disk outside the
compact object}. However this would not be enough energy to make a
powerful gamma-ray burst as proposed by Cen (1999). The radio
and x-ray afterglow that one might expect from the impact of such a
jet on the circumstellar medium was also not observed..

\vskip 0.2 in
\noindent
{\sl Still to be discovered} - It may be that, especially with common
GRBs, we have just seen the ``tip of the iceberg'' of a large range of
high energy phenomena powered by hyper-accreting, stellar mass black
holes. We already mentioned the possibility of a large population of
faint, soft bursts like GRB 980425. Other possibilities include very
long GRBs below the threshold of BATSE, ``orphan'' x-ray afterglows
from jet powered Type II supernovae, GRBs from the first explosions of
massive stars after recombination, and more. It is an exciting time.

\acknowledgements

The authors gratefully acknowledge helpful conversations on the
subjects of gamma-ray bursts with Chris Fryer, David Meier, Ewald
M\"uller, and Martin Rees. This work has been supported by NASA
(NAG5-8128 and MIT SC A292701), the NSF (AST 97-31569 and
INT-97-31569), and the Department of Energy ASCI Program
(W-7405-ENG-48), and by the Alexander von Humboldt-Stiftung (1065004).
We are also grateful to the Max Planck Institut f\"ur Astrophysik for
its hospitality while the final draft of this paper was prepared.

{}

\clearpage

\begin{deluxetable}{llllll}

\tablecaption{Explosion energy and fallback.
}
\tablehead{ 
\colhead{run} & 
\colhead{$\aP$} & 
\colhead{explosion energy\tablenotemark{a}} &
\colhead{KEP fallback} &
\colhead{PPM fallback} & 
\colhead{t$_{1/2}$\tablenotemark{b}} \\
\colhead{} & 
\colhead{} & 
\colhead{($\foe$)} &
\colhead{($\Msun$)}&
\colhead{($\Msun$)}&
\colhead{(s)} 
} 
\startdata
A01  & 0.025 & 0.255            & 3.71        & 3.63  & 100  \\         
A02  & 0.05  & 0.341            & 3.41        & 3.35  & 145  \\ 
A03  & 0.10  & 0.479            & 3.03        & 3.00  & 265  \\         
A04  & 0.15  & 0.595            & 2.85        & 2.64  & 450  \\         
A05  & 0.20  & 0.702            & 2.52        & 2.23  & 730  \\         
A06  & 0.25  & 0.805            & 1.96        & 1.73  & 1060  \\        
A07  & 0.30  & 0.906            & 1.39        & 1.22  & 1140  \\        
A08  & 0.35  & 1.007            & 0.91        & 0.83  & 890 \\          
A09  & 0.40  & 1.105            & 0.60        & 0.60  & 940  \\         
A10  & 0.42  & 1.152            & 0.53        & 0.53  & 1000  \\        
A11  & 0.44  & 1.207            & 0.48        & 0.47  & 1060  \\        
A12  & 0.50  & 1.326            & 0.24        & 0.33  & 1240  \\        
A13  & 0.60  & 1.507            & 0           & 0.18  & 1310 \\         
A14  & 0.70  & 1.682            & 0           & 9.8(-2)       & 970  \\ 
A15  & 0.80  & 1.850            & 0           & 6.2(-2)       & 610  \\ 
A16  & 0.95  & 2.092            & 0           & 3.7(-2)       & 440  \\ 
\hline      
A17\tablenotemark{c}                                                    
       & 0.025 & 0.316            & 3.37     & -  & - \\
\hline      
B01  & 0.20  & 0.314            & 5.2         & -     & -  \\ 
B02  & 0.30  & 0.415            & 4.9         & -     & -  \\ 
B03  & 0.40  & 0.513            & 4.6         & -     & -  \\ 
B04  & 0.60  & 0.726            & 3.1         & -     & -  \\ 
B05  & 0.70  & 0.835            & 2.1         & -     & -  \\ 
B06  & 0.75  & 0.913            & 1           & -     & -  \\ 
B07  & 0.80  & 0.960            & 0.06        & -     & -  \\ 
B09  & 0.95  & 1.143            & 0.05        & -     & -  \\ 
B10  & 1.00  & 1.200            & 0           & -     & -  \\ 
\tablenotetext{a}{Final kinetic energy of the ejecta at infinity.}
\tablenotetext{b}{Time scale to accrete half of total accreted mass.}
\tablenotetext{c}{Final piston radius of $\Ep{10}\,\cm$.}
\enddata
\lTab{Expl}
\end{deluxetable}

\clearpage

\begin{table*}
\vskip 8pt \centerline {Table 2. Explosion Characteristics at t = 400
s After Jet Initiation}
\begin{center}
\vskip 3pt
\begin{tabular}{ccccccc}
\hline 
      Name & $\epsilon$  & f$_{\rm P}$   & $\Delta$ M & E$_{tot}$ 
        & R($\theta > 10^\circ$) & R($\theta > 20^\circ$) \\[3pt] 
           &             &               & (M\sun) & (10$^{51}$\,ergs) & & \\
\hline
      J33 &  0.001  & 0.001 & 2.76 & 3.38 & 0.075 & 0.037 \\
      J32 &  0.001  & 0.01  & 2.69 & 3.23 & 0.102 & 0.047 \\
      J31 &  0.001  & 0.1   & 2.51 & 3.00 & 0.425 & 0.256 \\
      J22 &  0.01   & 0.01  & 1.72 & 19.91 & 0.429 & 0.230 \\
\hline
\end{tabular}
\end{center} 
\end{table*} 

\clearpage

\begin{figure}
\epsscale{0.60}
\plotone{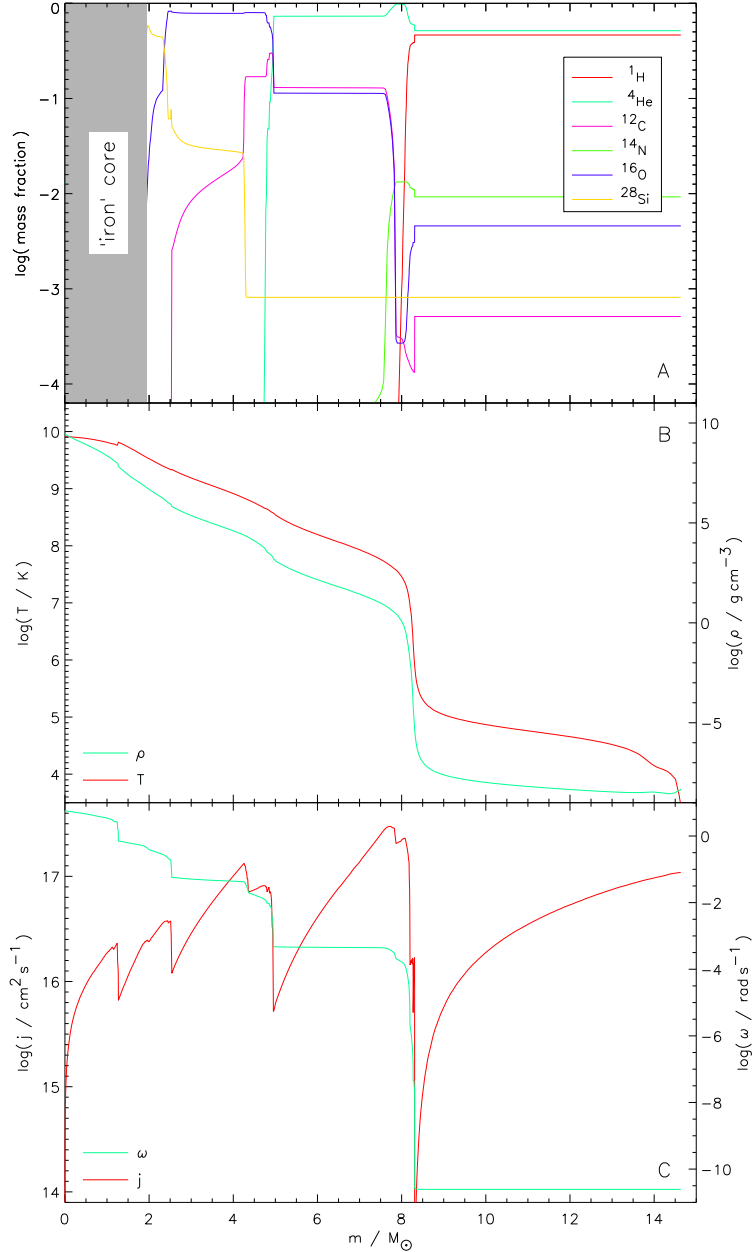}
\caption{
The presupernova model used for these studies, Model A, is derived
from calculations of Heger, Woosley, \& Langer (1999). This model was
evolved with ``restricted semiconvection'', including the effects of
rotation and an equatorial rotational velocity on the main sequence of
200 km s$^{-1}$. Panel A shows the composition of this initially 25
M\sun \ star, now reduced to 14.63 M\sun \ by mass loss. The iron core
was removed for this study and replaced by a piston
(\Fig{wilson}). Panel B shows the characteristic red supergiant
density and temperature structures of the presupernova star,
especially the cool low density hydrogen envelope outside 8.4
M\sun. The bottom panel shows the distribution of specific angular
momentum and angular velocity in the presupernova star. The angular
momentum in the equatorial plane is actually 50\% higher than this
angle-averaged value.
\lFig{presn}}
\end{figure}

\begin{figure}
\epsscale{0.8}
\plotone{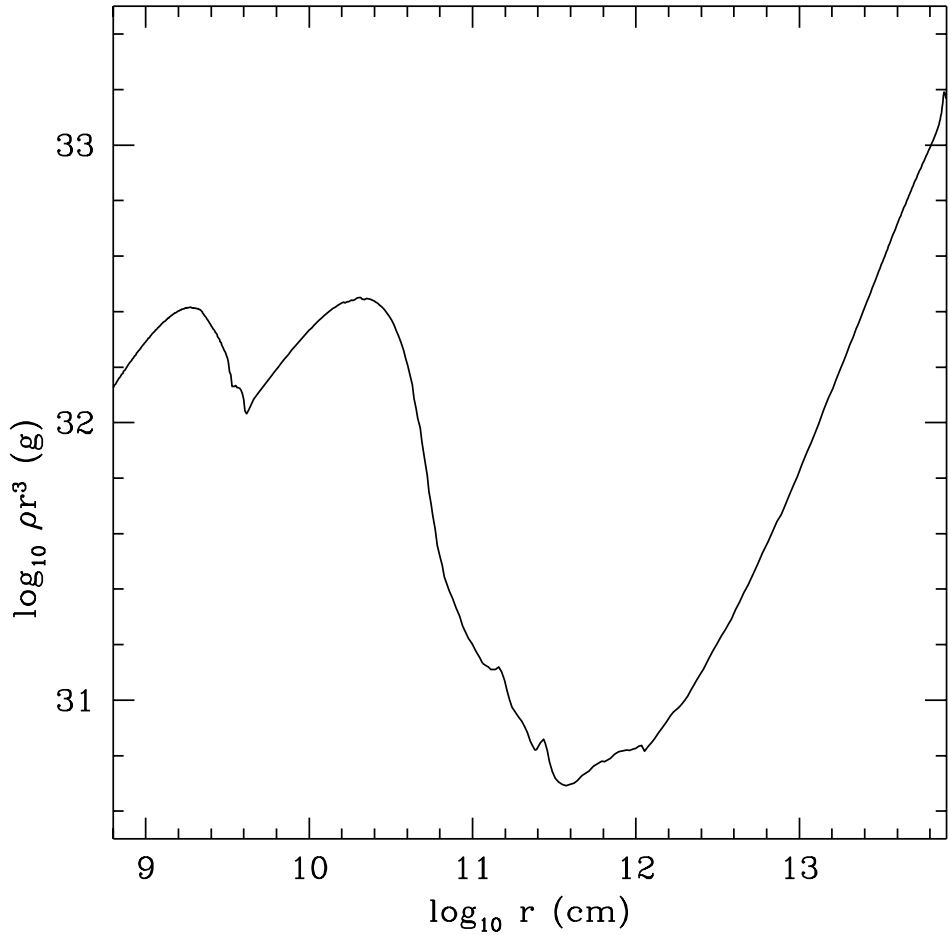}
\caption{
The distribution of the quantity $\rho r^3$ in the presupernova model
(\Fig{presn}). The supernova shock will speed up in regions of
decreasing $\rho r^3$ and slow in regions where $\rho r^3$
increases. The large increase outside 10$^{12}$ cm occurs in the
hydrogen envelope. The dip at log r = 9.6 and 10.8 are the edges of
the carbon-oxygen and helium cores.
\lFig{rhor3}}
\end{figure}

\clearpage
\begin{figure}
\epsscale{1.0}
\begin{center}
\begin{tabular}{c}
\epsfxsize = 5.5cm
\epsfbox[14 819 580 1618]{fig3a.ps}
\\
\epsfxsize = 5.5cm
\epsfbox[14 819 580 1618]{fig3b.ps}
\end{tabular}
\caption{
Radial history of Lagrangian mass shells in the KEPLER calculation of
Model A01 (\Tab{Expl}). Dark lines represent mass shells of 0.5 M\sun,
faint lines, 0.05 M\sun. The piston (bottom-most dark
line) moves out, launching the shock, and comes to rest, after 19.5 s,
at 10$^9$ cm. The location of the supernova shock is visible as the
outer boundary of the dark concentration of curved lines. All material
on the grid moves outwards until 60 s after core bounce, when some begins
to to fallback and come to rest on the piston. After about 150 s, an
accretion shock has developed whose location is an artifact of the
stationary piston. In the lower panel the supernova shock continues to
the surface of the star, but a total of 3.71 M\sun \ eventually falls
back to rest on the piston. The slope of the shock location gives its
speed which decelerates in regions of increasing $\rho r^3$. This
model was linked to the PPM code 100 s after material had begun to
fallback, but before any appreciable accumulation on the piston.
\lFig{wilson}}
\end{center}
\end{figure}

\clearpage

\begin{figure}
\epsscale{1.0}
\plotone{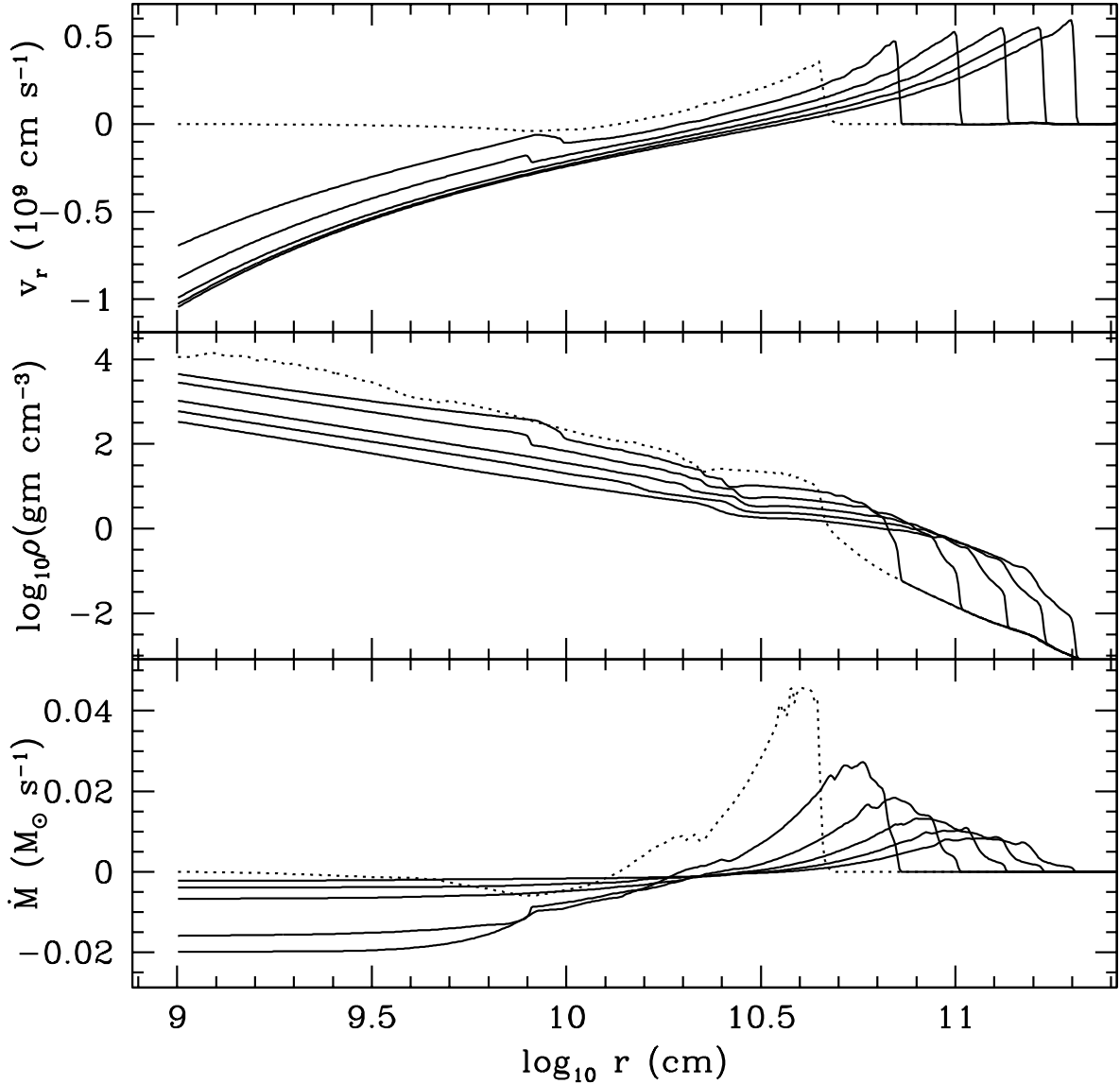}
\caption{ 
Propagation of the shock shown in \Fig{wilson} following the remapping
of the problem into the one-dimensional PPM code at 100 s (dotted
line). The solid lines show the velocity (10$^9$ cm s$^{-1}$), log
density (g cm$^{-3}$), and accretion rate (M\sun \ s$^{-1}$) as a
function of radius at intervals 50, 100, 150, 200, and 250 s after the
remapping.  During the first roughly 20 s the motion of matter is
still influenced by memory of the KEPLER piston at 10$^9$ cm and is
nearly stationary at small radii. Though the residual support of this
material is artificial, only a small amount of mass accretes at this
time, which may be thought of as the interval during which the loss of
pressure at the origin due to black hole formation propagates to
10$^9$ cm. By 250 s after the remapping, (last solid line shown) the
accretion rate has declined greatly (\Fig{mdot}).
\lFig{waves}}
\end{figure}


\clearpage

\begin{figure}
\epsscale{1.0}
\plotone{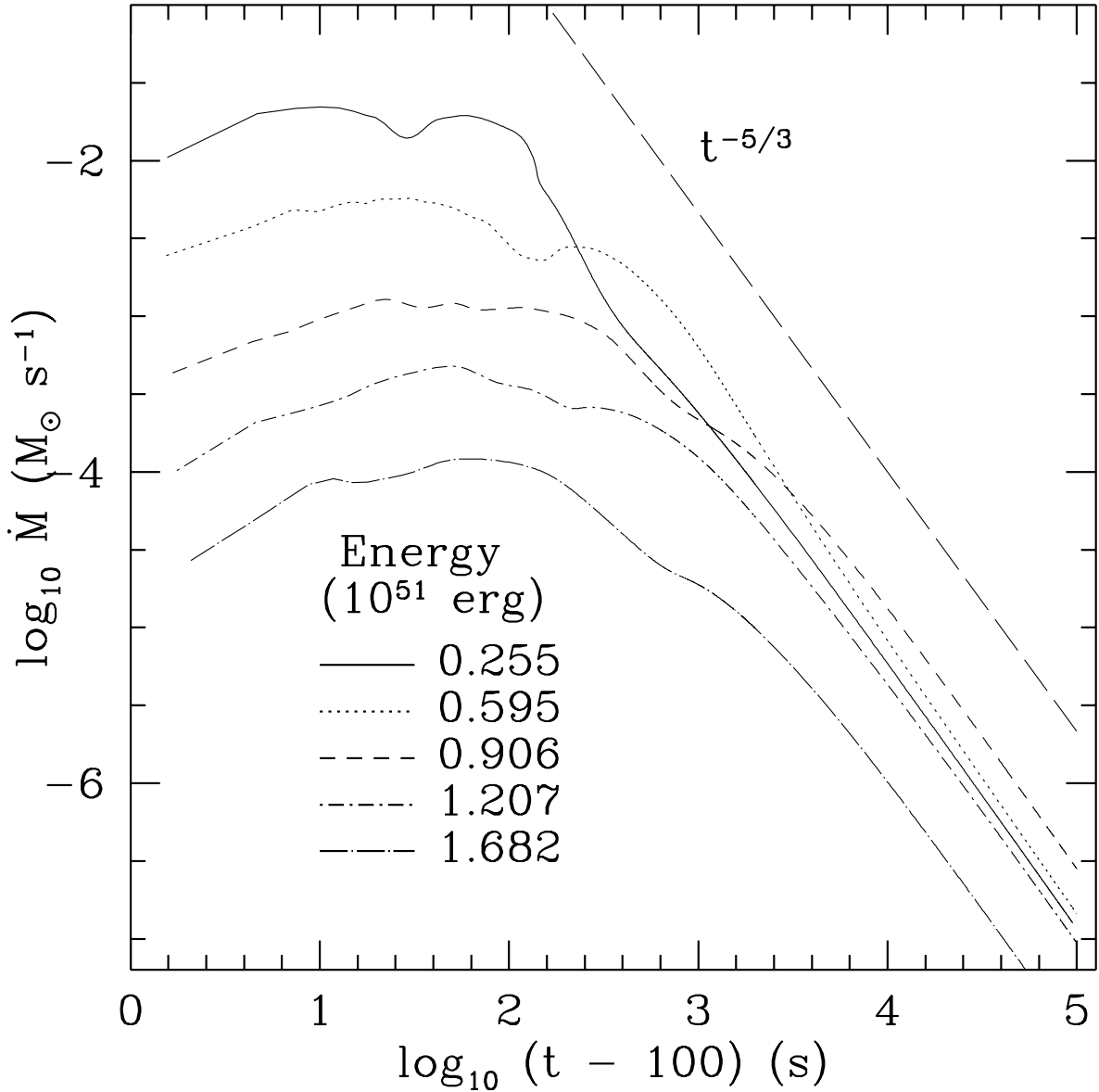}
\caption{ 
Accretion rates for fallback in five different explosions: A01
(solid), A04 (dotted), A07 (dashed), A11 (short dash-dot) A14 (long
dash-dot) (Table 1). Calculations were carried out using a
one-dimensional version of the PROMETHEUS PPM hydrodynamics code.
Time is measured from the collapse of the iron core and accretion rate
plotted as a function of the time since the calculation was linked to
the PPM code at 100 s. At late times the accretion rate follows the
t$^{-5/3}$ scaling suggested by Chevalier (1989).
\lFig{mdot}}
\end{figure}


\clearpage

\begin{figure}
\epsscale{1.0}
\plotone{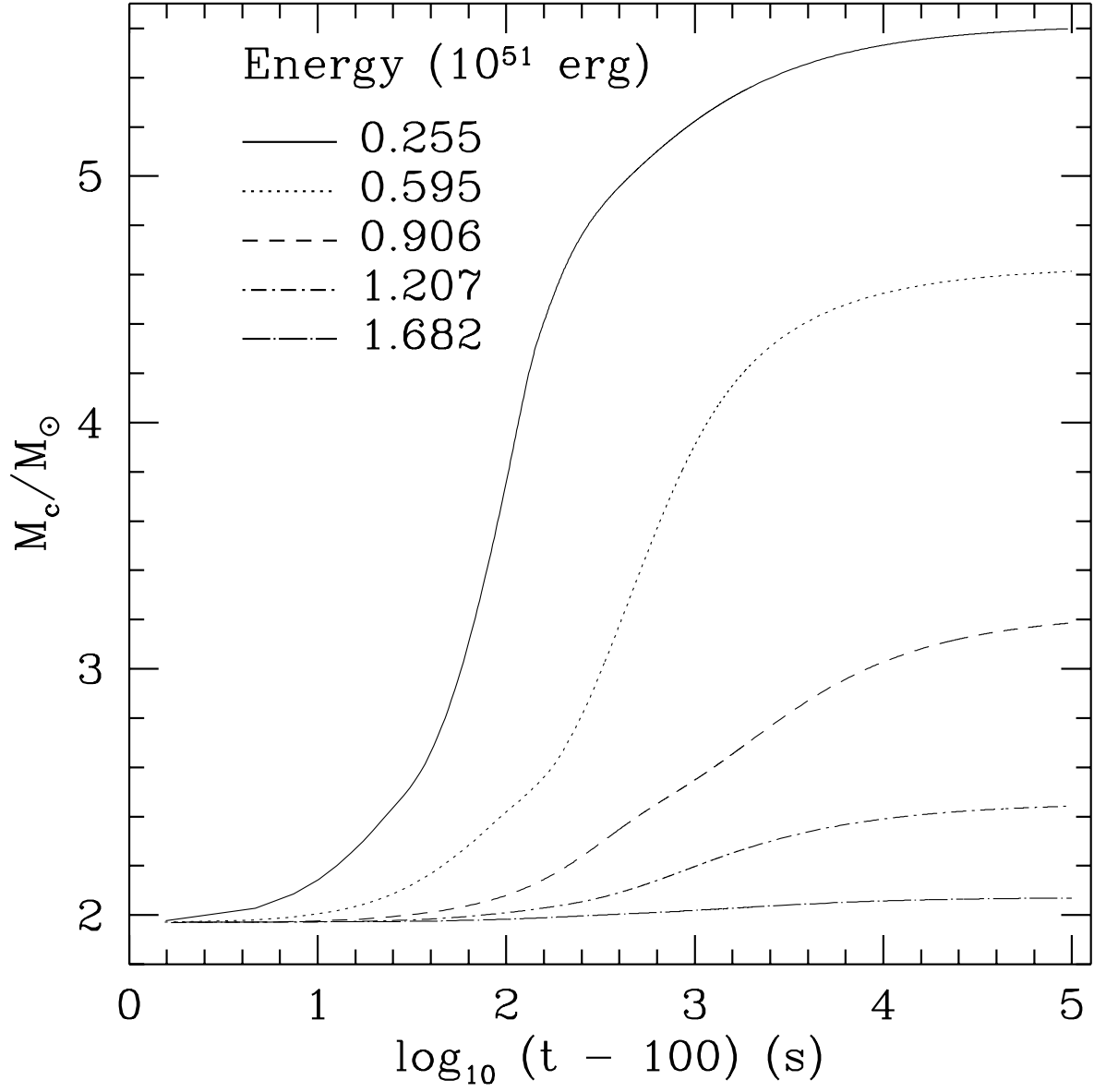}
\caption{ 
Accumulated mass as a function of time for the same calculations
shown in \Fig{mdot}.
\lFig{totm}}
\end{figure}

\clearpage

%
\begin{figure}
\epsscale{1.0}
\begin{center}
\begin{tabular}{cc}
\epsfxsize = 8.0cm
\epsfbox{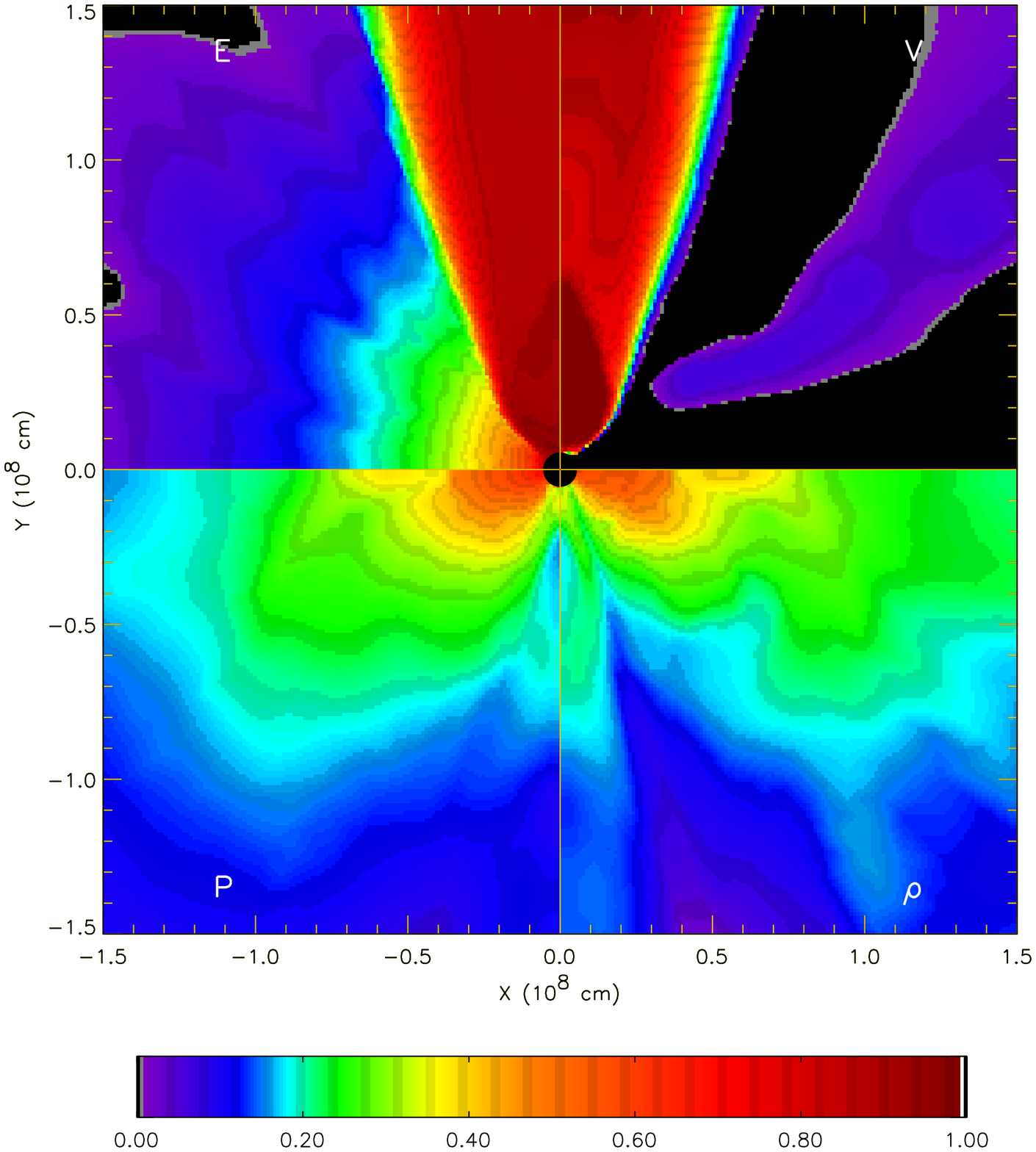}
&
\epsfxsize = 8.0cm
\epsfbox{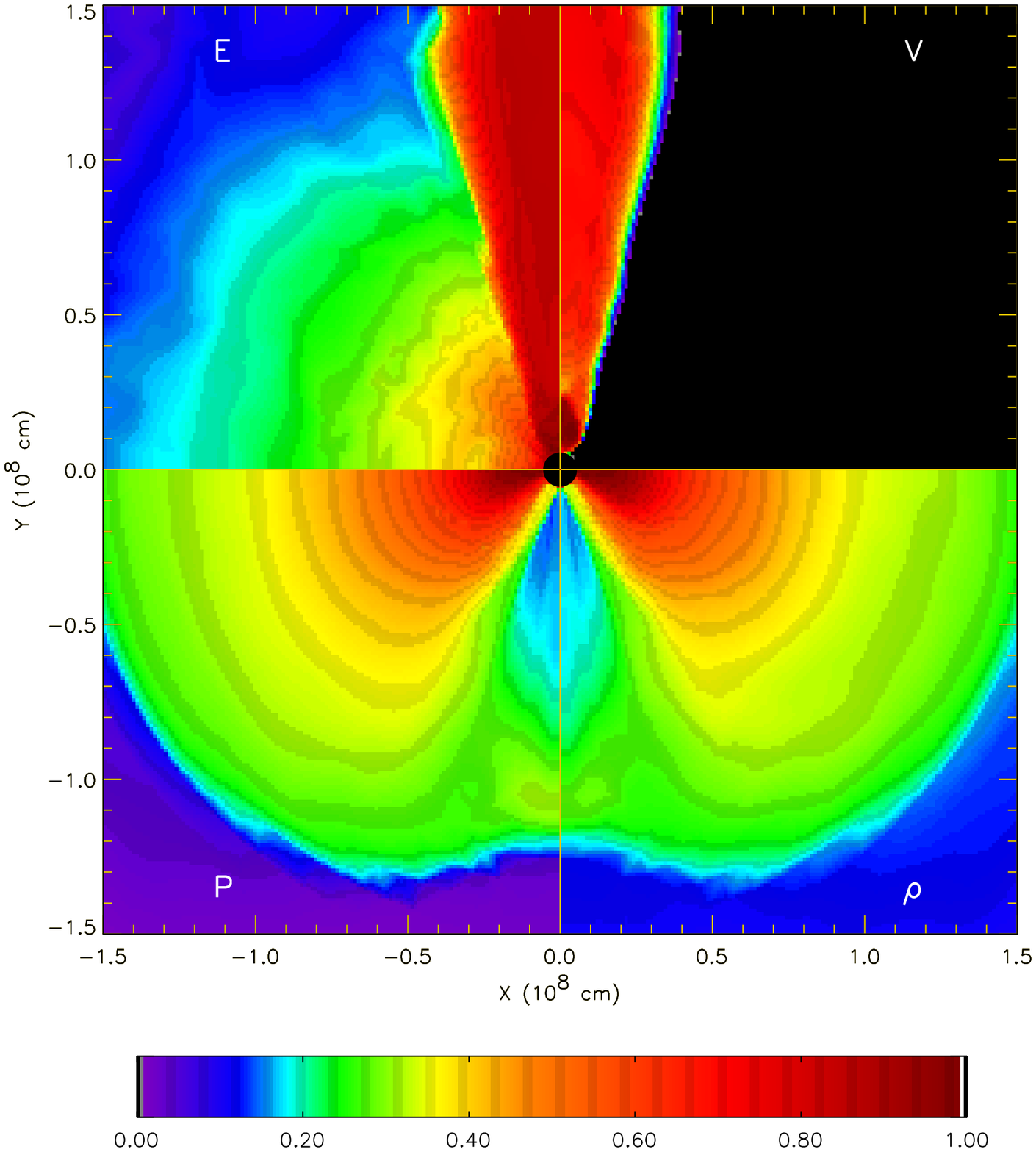} 
\end{tabular}
\caption{
Hydrodynamical focusing of jets initiated in the collapsar disks of
MW99 is shown here for two values of the disk viscosity parameter,
$\alpha = 0.1$ (left) and $5 \times 10^{-4}$ (right).  Each jet was
injected at $5 \times 10^6$ cm with an energy deposition rate of $1.8
\times 10^{51}$ erg s$^{-1}$, roughly equally in internal and kinetic
energy, and an opening half-angle of 10 degrees. The figures show the
situation 0.6 s after the jets are turned on.  The bottom two
panels of each figure show the logarithms of the pressure (left) and
density (right) for the disk before jet initiation. The mass of the
disk is approximately two orders of magnitude smaller in the high
viscosity case. The top two panels show the logarithm of total energy
density (left) and the velocity (right) 0.6 s after the start of
jet injection. The minimum and maximum values for internal energy are
log $E$ (erg g$^{-1}$) = 18.5 and 20.3, for velocity, $v$ (cm s$^{-1}$) =
$5.0 \times 10^8$ and $1.5 \times 10^{10}$, for density, log $\rho$ (g
cm$^{-3}$)= 5 and 11.5, and for pressure, log $P$ (dyne cm$^{-2}$) = 23
and 30.1. The color bar indicates the logarithmic interpolation
between these two extrema.  For example, the yellow-orange region of
the density plot represents $\log \rho = 5 + 0.40(11.5 - 5) = 7.6$.
The jet is clearly less well collimated in the high viscosity case,
presumably because of the smaller mass of the disk and the weaker
pressure gradients confining the flow to the polar region.  Note also
the presence of ``plumes'' of high outward velocity at about 45
degrees in the plot for the high viscosity case. See MW99 for
discussion.
\lFig{jetinit}}
\end{center}
\end{figure}

\clearpage

\begin{figure}[t]
\epsscale{1.0}
<\psfig{file=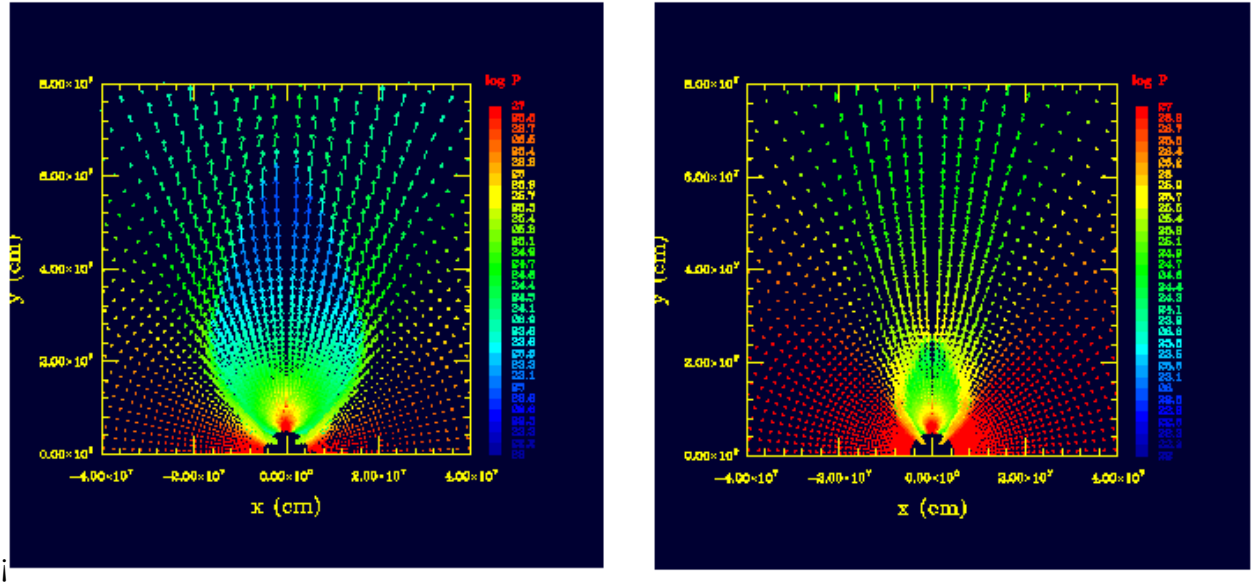,width=\textwidth}
\caption{
Further detail of the initial hydrodynamical focusing of the initial
jets in the high (left) and low (right) disk viscosity cases (see
\Fig{jetinit}). The pressure ranges from 10$^{22}$ dyne cm$^{-2}$
(dark blue) to 10$^{27}$ dyne cm$^{-2}$ (red) and the region plotted
is the inner 400 km ($x$) by 800 km ($y$). The longest velocity vectors
represent a speed of $1.5 \times 10^{10}$ cm s$^{-1}$. Both plots are
at a time 0.6 s after jet initiation.  The lower density in the high
viscosity case results in the jet being less well collimated. It
expands laterally with approximately the sound speed, which for equal
internal and kinetic energies is comparable to its radial speed. A jet
with a higher fraction of internal energy would spread even more,
especially if the disk mass were further reduced. The jet in the low
viscosity calculation also propagates in a lower density medium and
moves further before a shock develops.
\lFig{flamejet}}
\end{figure}

\clearpage

\begin{figure}[t]
\psfig{file=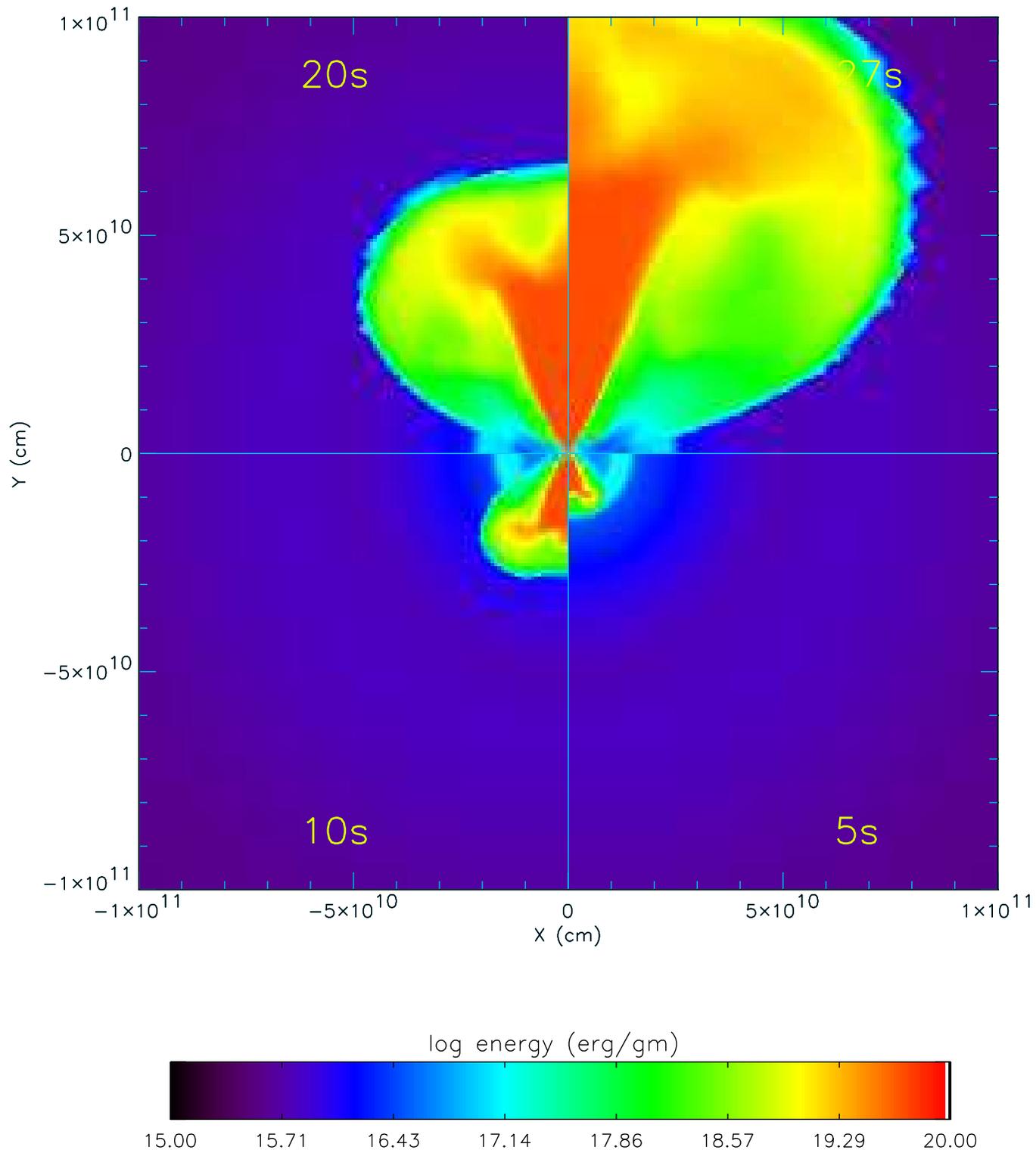, width=\textwidth}
\caption{
The specific energy density (internal plus kinetic) of the jet and
explosion is shown at times of 5, 10, 20 and 27 s after initiation of
the jet in Model J22 (Table 2). The passage of the jet initiates a
shock that propagates to lower latitudes, eventually exploding the
entire star.  The original supernova shock can be seen as a pale blue
circle at a radius of about $2\times 10^{10}$ cm.
\lFig{ecomp}}
\end{figure}


\clearpage


\begin{figure}[t]
\psfig{file=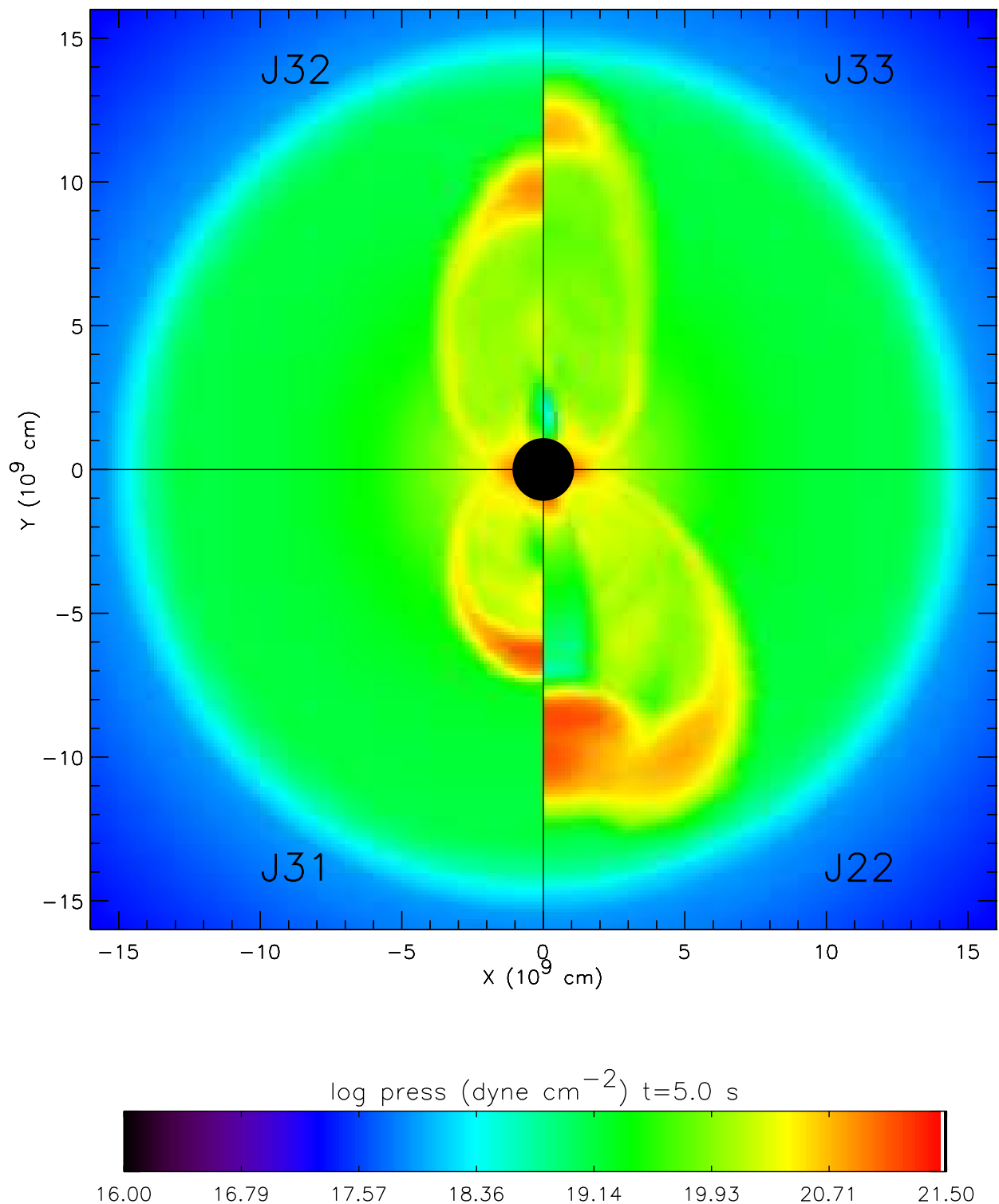,width=\textwidth}
\caption{
Pressure in the jet and surrounding star at 5.0 s after the initiation
of the jet in four different models (Table 2). For the J3n series,
higher pressure clearly leads to greater jet divergence, more mass
swept up, and slower propagation. Model J22 had a higher jet energy
than the other models. The edge of the helium core is at $5 \times
10^{10}$ cm (X$_{\rm H}$ = 0.01; $\rho$ = 0.6 g cm$^{-3}$), but the
presupernova density begins to decrease rapidly below 100 g cm$^{-3}$
outside of $1.4 \times 10^{10}$ cm (the green disk in the figure).
\lFig{press}}
\end{figure}


\clearpage


\begin{figure}[t]
\psfig{file=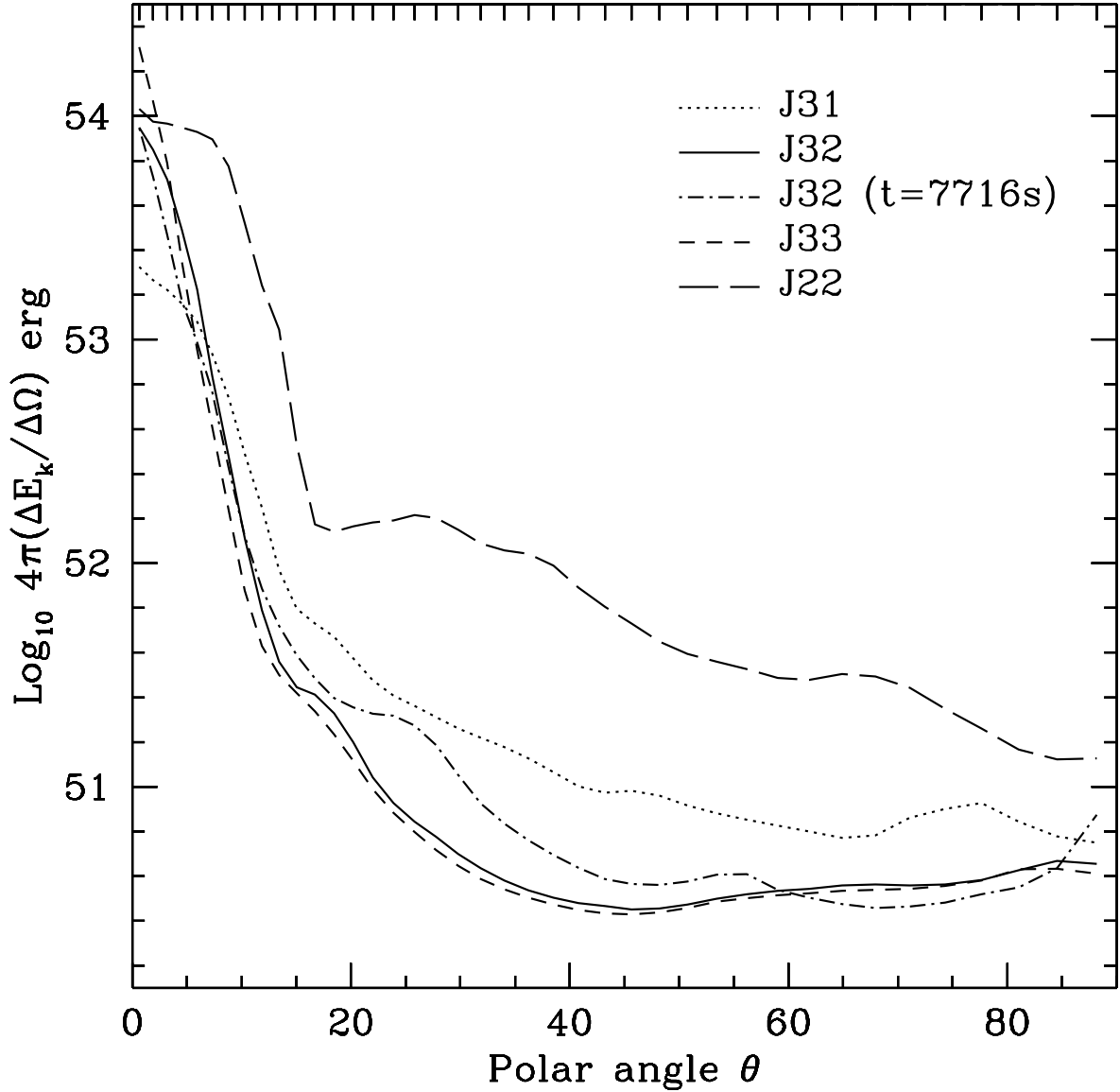,width=\textwidth}
\caption{
The ``equivalent isotropic kinetic energy'' as a function of polar
angle for four models having variable energy efficiency factors and
internal pressures (Table 2). Except for J32, the calculations are all
sampled at 400 s after the initiation of the jet (500 s
post-bounce). At this point the jets have exited the helium core and
are moving through the hydrogen envelope. Model J32 is shown at two
times, once at 400 s and later, at 7716 s, as the jet reaches the
surface of the star at $8 \times 10^{13}$cm (dash-dot line).
The collimation of Model J32 is further improved by its passage
through the hydrogen envelope.  Note that the degree of collimation is
strongly dependent upon $f_{\rm P}$.  Equivalent isotropic kinetic
energy is defined as the integral from the center to surface of the
star of its kinetic energy in the solid angle subtended by $\theta$
and $\theta + \Delta\theta$ divided by the solid angle,
2$\pi(\cos\theta - \cos(\theta + \Delta\theta))$ and multiplied by
4$\pi$.  The injected energy at the base of the jet would be a flat
line out to ten degrees with a value equal to $66 \, \epsilon \Delta
{\rm M} c^2$ with $\Delta {\rm M}$ in Table 1 and 66 = (1 -
$\cos(10^\circ))^{-1}$.  Tick marks along the top axis give the
angular zoning of the two dimensional code.
\lFig{equiso}}
\end{figure}


\clearpage

\begin{figure}
\epsscale{1.0}
\begin{center}
\begin{tabular}{cc}
\epsfxsize = 8.0cm
\epsfbox{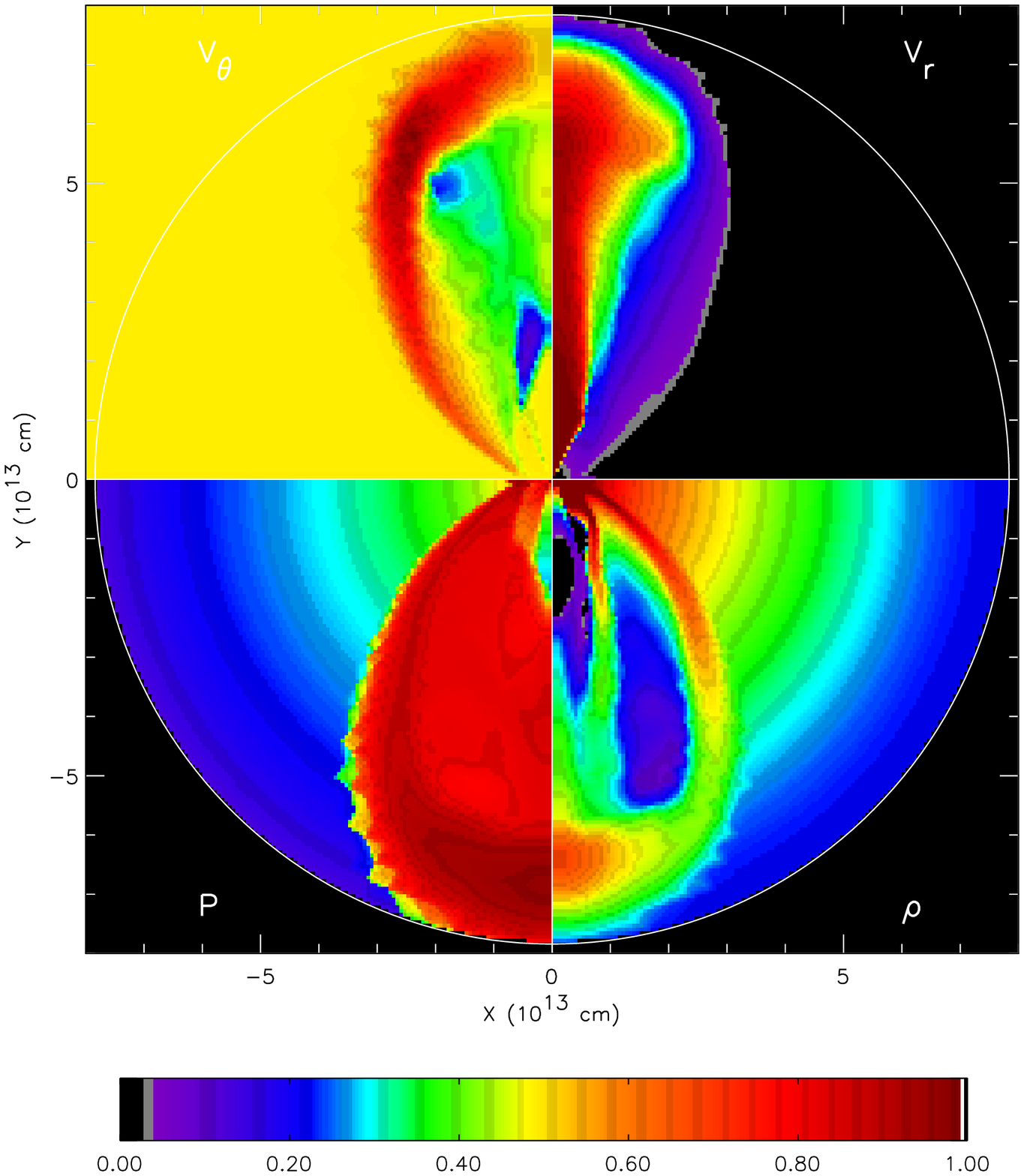}
&
\epsfxsize = 8.0cm
\epsfbox{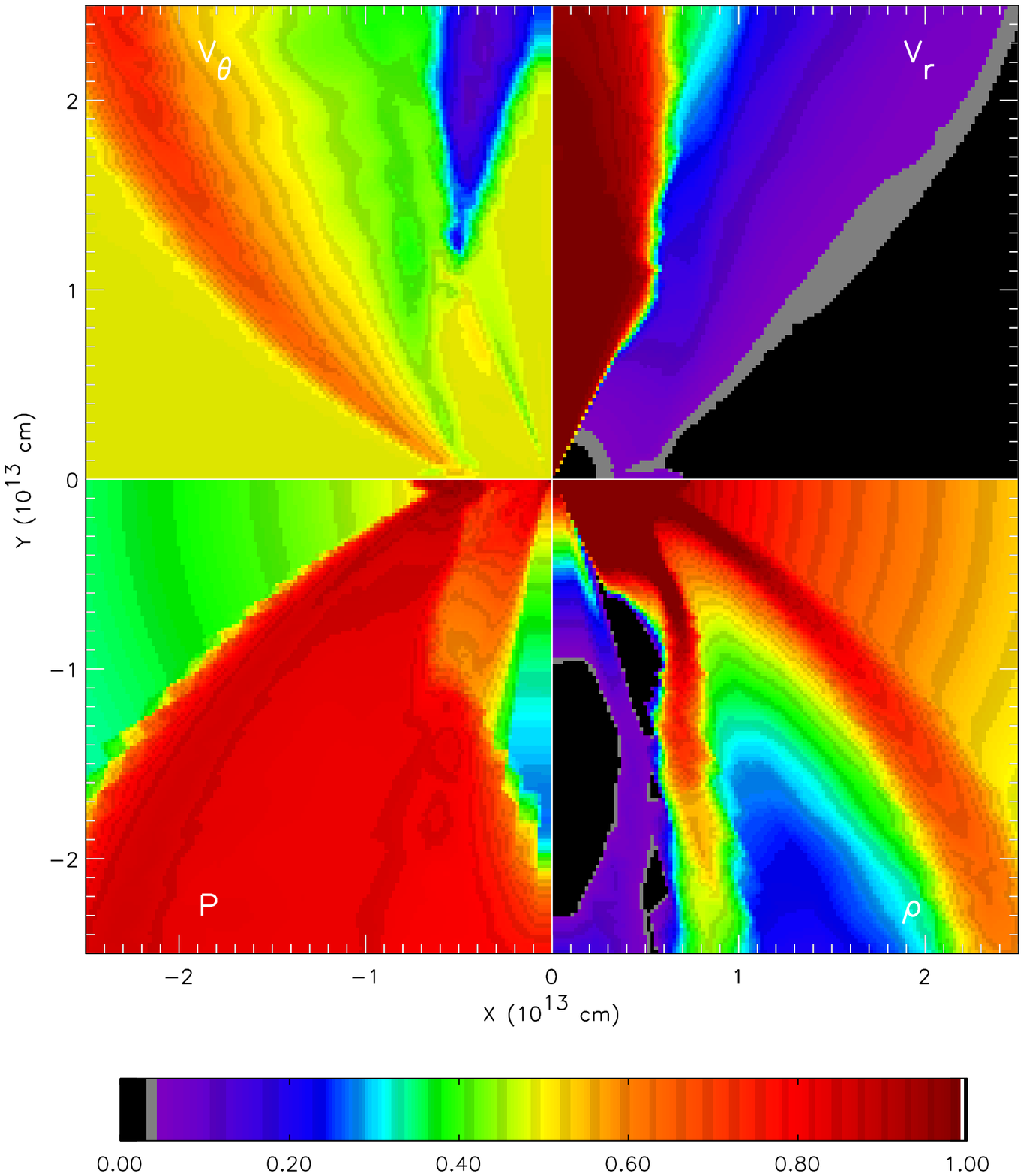} 
\end{tabular}
\caption{
The structure of Model J32 as the jet nears the surface 7820 s after
core collapse.  The total explosion energy at this time is $4.1 \times
10^{51}$ ergs at this time, probably a good approximation to the final
value.  The theta velocity, radial velocity and logarithms of density
and pressure are given with the minimum and maximum values for
$v_{\theta}$ are $-1.5 \times 10^9$ and $1.5 \times 10^9$, for radial
velocity, $v_r$, -6.7 $\times 10^7$ and 1.0 $\times 10^{10}$, for
density, log$_{10} \rho$, -9 and -6.8, and for pressure, log$_{10}$ P,
2.5 and 10.9, all in cgs units.  The colors indicate the interpolation
scale between minimum and maximum (see \Fig{jetinit}).  Positive
$v_{\theta}$ is motion away from the polar axis ($\theta = 0$) along
an arc of constant radius.  The $v_{\theta}$ plot shows the expansion
of the high pressure bubble blown by the jet sweeping around the star
(red region) but also an inner region of collimation (blue, purple,
and green).  At $r = 7 \times 10^{13}$ cm the $x$-component of the
velocity of the expansion shock is $v_x = 2.0 \times 10^9$ cm s$^{-1}$
(compared to a sound speed of only $3 \times 10^6$ cm s$^{-1}$), while
the $y$-component is $v_y = 9.6 \times 10^{9}$ cm s$^{-1}$ resulting in
an aspect ratio for the bubble between 0.2 and 0.3.  The velocity and
pressure plots show a collimation of the jet flow near $1 \times
10^{13}$ cm, well into the hydrogen envelope of the star, and
evacuation of a low density column by the jet. Not however, the plug of
high density material being shoved along by the jet.
\lFig{breakout}}
\end{center}
\end{figure}


\clearpage


\begin{figure}
\epsscale{1.0}
\plotone{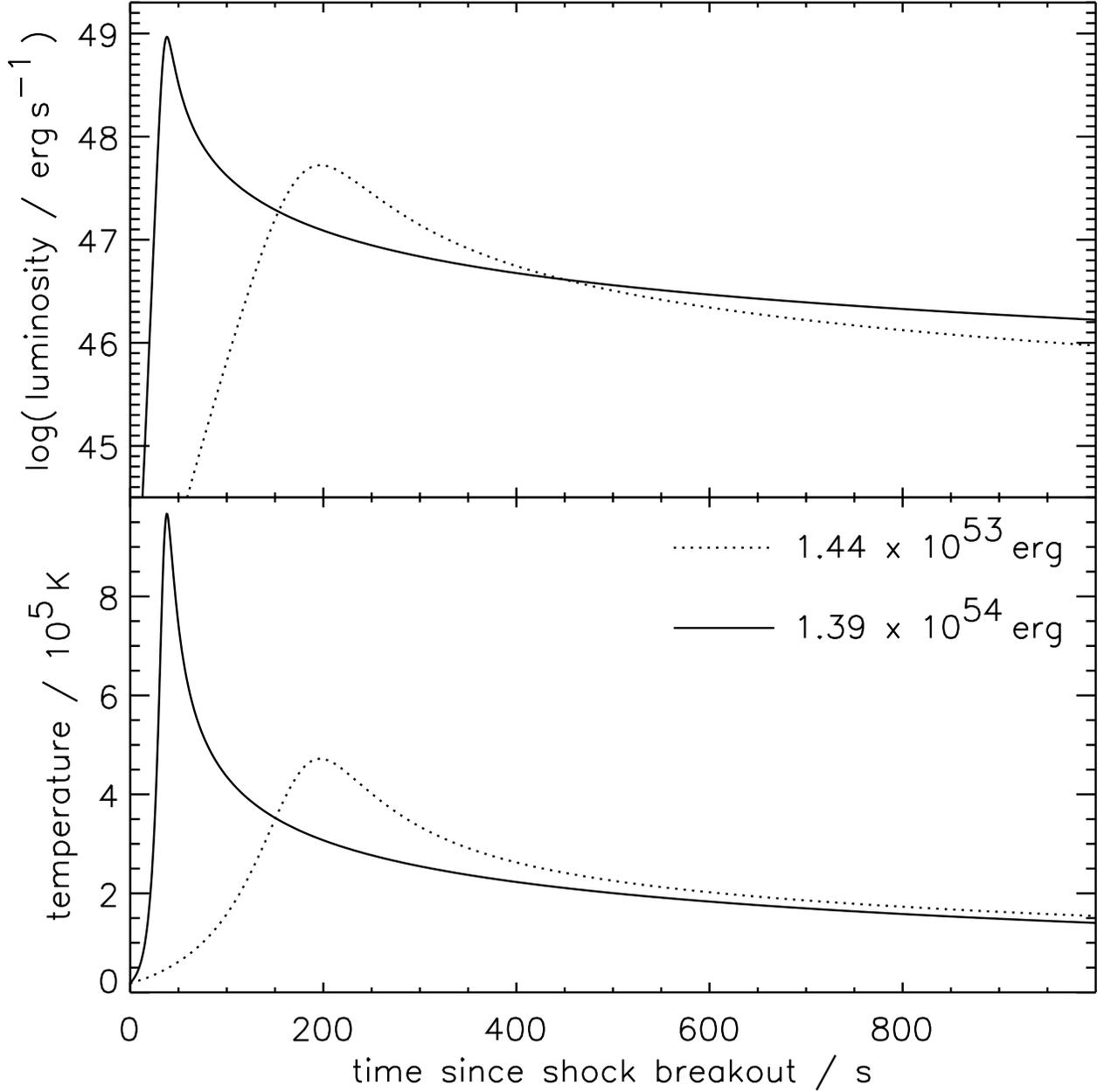}
\caption{
Shock breakout in spherically symmetric explosions for Model A for the
KEPLER calculation of two cases of a $1.44 \times 10^{53}$ erg
explosion (dotted line) and a $1.39 \times 10^{54}$erg explosion
(solid line) as a function of time since the onset of shock breakout.
The upper panel gives the total luminosity and the lower panel the
{\it effective} temperature, which probably underestimates the color
temperature, T$_{\rm c}$ by about a factor of two to three. The FWHM
of the luminosity curves are 12.5 and 94 s for low and high energy
respectively. The widths of the temperature curves are 56 and 350 s.
\lFig{softx}}
\end{figure}

\clearpage

%
\begin{figure}
\epsscale{1.0}
\begin{center}
\begin{tabular}{cc}
\epsfxsize = 8.0cm
\epsfbox{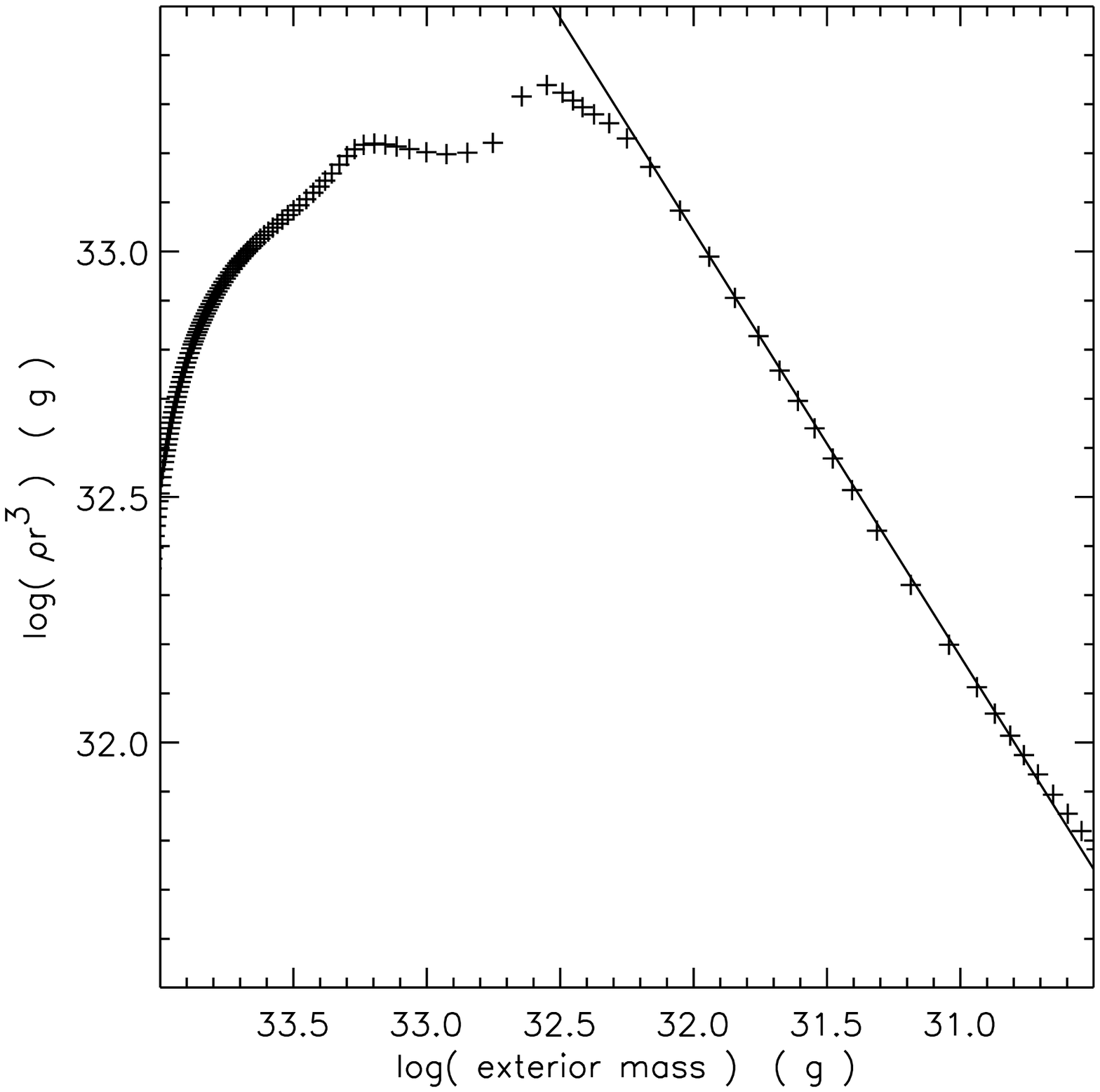}
&
\epsfxsize = 8.0cm
\epsfbox{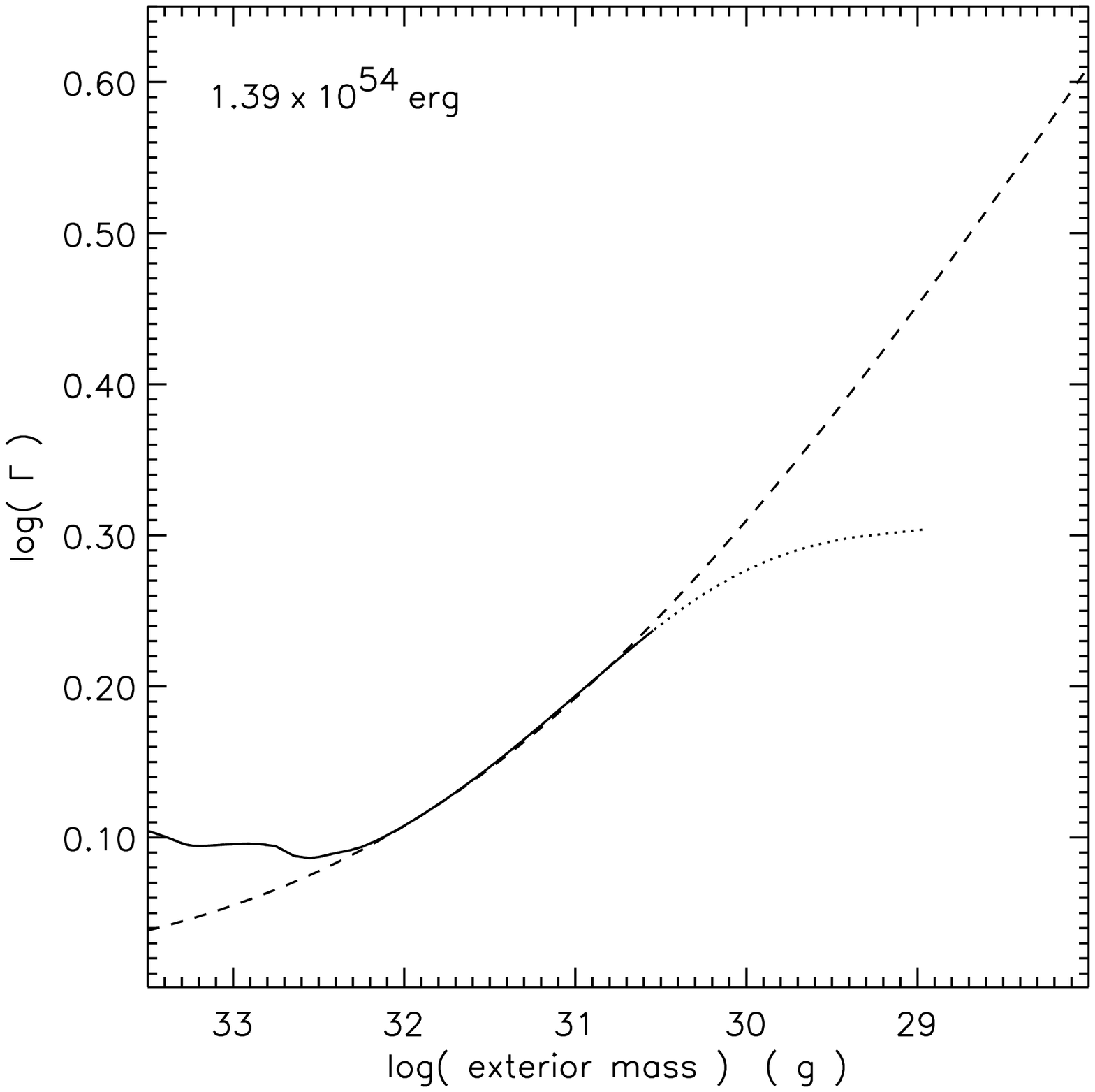} 
\end{tabular}
\caption{
Density structure near the surface of the red supergiant (Model A)
(left panel) and extrapolated Lorentz factor for the breakout of a
shock of energy $1.39 \times 10^{54}$ erg (right panel). Crosses show
the product of density times r$^3$ for the hydrogen envelope and zones
near the surface. A power law $\rho r^3 \propto (\Delta m)^{0.86}$
fits the surface of the Kepler model well. Using this density
structure and the near constancy of $\Gamma \beta (\rho r^3)^{1/5}
\approx 3.25 \times 10^6$ (g$^{1/5}$), one can extrapolate (dashed
line) to obtain the relativistic $\Gamma$ for mass zones outside of
10$^{30.5}$ g. However the curve for $\Gamma$ is not valid for
external masses less than 10$^{29.3}$ g (see text). The maximum
$\Gamma$ for this shock energy is thus $\Gamma \approx 2.5$.  The
curved dotted line at masses below 10$^{30.7}$ shows the results one
obtains using the actual density strcture in KEPLER for optically thin
zones rather than the extrapolation of $\rho r^3$ in the left panel.
\lFig{gamma}}
\end{center}
\end{figure}

\end{document}